\documentclass[11pt]{article}
\usepackage{amsmath,amssymb,amscd,amsthm,graphicx,float,fullpage,mathrsfs,centernot,hyperref,stackrel,subcaption}
\usepackage[export]{adjustbox}

\newcommand{\X}{\mathbf{X}}

\usepackage{color}

\makeatletter
\newcommand*{\rom}[1]{\expandafter\@slowromancap\romannumeral #1@}
\makeatother
\usepackage{graphicx}
\usepackage{graphics}
\graphicspath{{plot/}}
\usepackage{subcaption}
\usepackage[top=1in, bottom=1in, left=1in, right=1in]{geometry}
\usepackage{float}
\restylefloat{table}
\usepackage{xcolor,authblk}
\usepackage{titlesec}
\usepackage{bm}
\titleformat{\section}
  {\normalfont\Large\bfseries}{\S\thesection}{1em}{}
\usepackage{setspace}
\usepackage{changes}
\definechangesauthor[color=violet]{EEK}
\setauthormarkup{}

\title{On variance estimation for generalizing from a trial to a target population}

\author[1]{Ziyue Chen}
\author[1]{Eloise Kaizar}

\affil[1]{Department of Statistics, The Ohio State University, Columbus, OH, USA}

\begin{document}
\maketitle
\singlespacing

\begin{abstract}
Randomized controlled trials (RCTs) provide strong internal validity compared with observational studies. However, selection bias threatens the external validity of randomized trials. Thus, RCT results may not apply to either broad public policy populations or narrow populations, such as specific insurance pools. Some researchers use propensity scores (PSs) to generalize results from an RCT to a target population. In this scenario, a PS is defined as the probability of participating in the trial conditioning on observed covariates. We study a model-free inverse probability weighted estimator (IPWE) of the average treatment effect in a target population with data from a randomized trial. We present variance estimators and compare the performance of our method with that of model-based approaches. We examine the robustness of the model-free estimators to heterogeneous treatment effects.

\end{abstract}
\section{Introduction}\label{sec:intro}


Because some researchers believe that the analytical results from randomized controlled trials (RCTs) may not reflect the true average treatment effects that would be seen if the intervention were implemented in a different target population (e.g., \cite{altman2001,cole10,stuart11, kennedy2015}), several methods have been proposed to account for differences between RCT and target populations. We focus on reweighting methods such as those proposed by Cole and Stuart 2010 and Zhang et al 2015 \cite{cole10,zhang15}. Although these proposed reweighted average effect estimates have been promising, their characteristics have not yet been fully developed. In this paper, we lay out a new framework for considering the statistical properties of reweighted estimators that differentiates between a particular finite target population of interest and a more general population similar to an observational dataset. We examine the statistical properties of several reweighted average effect estimators under different estimation scenarios within this framework. Further, our framework clarifies the potential to improve variance estimation under various scenarios. We propose a new large sample variance estimator and detail two resampling -based variance estimators for several reweighted average effect estimators and examine their performance relative to existing variance estimation methods. 

Reweighting methods attempt to extract relative benefits of both RCT and observational study designs. The necessary conditions for making proper causal inference about treatment effects are guaranteed in expectation by the randomization in RCTs.  However, typical RCT participant recruitment leads some to question the validity of RCT-based results for target populations \cite{greenhouse08,imai08,cole10,stuart11,kennedy2015,zhang15}. Conversely, many observational studies are designed to reflect populations of interest, either via census (as in administrative data) or by relying on probability sampling. But these data either do not contain information about treatment or outcome at all, or possibly suffer from hidden confounding due to participants selecting their own treatments. Neither design on its own is sufficient for sound average treatment effect estimation in a target population.

Differences in the distribution of treatment effect modifiers between randomized trials and target populations can lead to generalizability bias if extrapolations ignore such differences (e.g., \cite{greenhouse08, stuart11, nie2013, hartman2015, kaizar2011}). The difference may be due to many possible design choices or other differences in study characteristics, including the time and location of the study, or clinical characteristics of the participants \cite{hernan2008, nie2013,hartman2015,zhang15}. In their seminal paper, Hernan, \textit{et al.} (2008)\cite{hernan2008} demonstrated the potentially crucial impact of differences in design (and resulting analyses) by showing that if one analyzes an observational study in a manner consistent with the design of a randomized trial, then the previously disparate estimates may also become consistent. 

While many design choices may be important, Stuart \textit{et al.} (2011) \cite{stuart11} focus on differences in patient characteristics across studies and propose a measure to quantify this difference that is based on the propensity of a participant to be recruited into a trial based on his or her observed baseline covariates.  Conceptually, this propensity is identical to a propensity score employed in traditional causal inference, defined as the conditional probability that a participant receives the treatment of interest \cite{rosenbaum83}. As such, we follow the lead of Stuart \textit{et al.} (2011) \cite{stuart11} to call the conditional probability of recruitment a {\em propensity score} (PS).  Stuart \textit{et al.} (2011) \cite{stuart11}go on to propose that if the difference between the two studies is substantial, then PSs can be used to adjust the difference in distributions between trials and target population to reduce or eliminate generalizability bias. Cole and Stuart (2010) \cite{stuart10} illustrate such a PS reweighting method, showing how standard survey sampling software designed for survival analysis can be used to estimate generalized average treatment effects.

Several authors have implemented reweighting or calibration schemes in different but similar contexts. For example, Hartman \text{et al.} (2015) \cite{hartman2015}calibrate an RCT to match a target population that received treatment already and thus estimate the population average treatment effect on the treated. Nie \textit{et al.} (2013) \cite{nie2013}, reweight the likelihood function of a noninferiority trial (or a historical well-controlled trial) to a similar population whose subjects can be represented by the trial. Weiss and Varadhan (2014) \cite{weiss2014}reweight a RCT to a target population whose treatment modifier distributions are not overlapping, through an additional larger intermediate trial. 

Corresponding to these reweighted estimators, various methods have been used to estimate their variances, including those based on the asymptotic theory of M-estimation for inverse probability weighted estimators (IPWEs) \cite{lunceford04,zhang15}, linearization approximations for survey sampling estimators (e.g.,\cite{lohr2009}), and bootstrap methods (e.g., \cite{rao85, efron79, nie2013, hartman2015}). In addition, one might consider variance estimation methods developed for weighted least square estimators, although the theory behind this method treats the weights fundamentally differently than the other methods.  Whereas weights used in methods based on traditional probability sampling and M-estimation are seen as how many people in the target population the subjects in the trial represent, weights used in methods for weighted regression are treated as the inverse variance of random errors.  

Different variance estimation methods also treat the calibration variables and weight models differently. Standard linearization and other probability sampling methods,  for which many statistical programming tools have built-in functions to estimate the variances, are easy to use but do not account for the variability of estimated weights and treat the target population as known and fixed. The M-estimation method used to obtain asymptotic variance captures the variability of estimated weights, covariates, outcomes, and experiment design elements, but are typically not readily available in software. Bootstrap methods can often be designed to match the stochastic process of the study, are straightforward to implement and can provide consistent estimators of variance with a large number of resamples, but the algorithm is computationally expensive and random resampling gives such estimates limited reproducability. To our knowledge, these various approaches to variance estimation have not been compared in the generalizability setting, nor has a large-sample variance estimation for the generalization IPWE been developed.

In addition to reweighting approaches, we note that other authors have proposed alternative methods for correcting generalizability bias.  For example, Greenland (2005)'s \cite{greenland2005} multiple bias models and Eddy \textit{et al.} (1992)'s \cite{eddy1989} confidence profile method are general approaches typically applied to adjust naive estimators for biases due to study design choices via parametric models.  These general approaches could, in theory, be used for the generalizability bias problem. In addition, Kaizar (2011) \cite{kaizar2011} and Weiss and Varadhan (2014) \cite{weiss2014} have both proposed methods for correcting generalizability bias when observational studies are available that include both comparative treatments and outcome measures.  In practice, such observational data is rarely available.  In addition to these limitations, we choose to focus on reweighting methods because there is natural interest among medical researchers in reweighting methods similar to the well-understood propensity-based methods used for causal inference. 

In particular, we develop IPWE for generalizing RCT results to a target population and propose a variance estimation method for developed IPWE. Moreover, we examine the properties of model-based estimators and IPW-based estimators and employ various variance estimation methods. The performance of the average treatment effect estimators is impacted by the subjects characteristics of RCTs and the target populations. Different variance estimation approaches may be appropriate for different goals, eg. estimating for a fixed finite population, or for an infinite population. We also examine the limitations of IPWE and the variance estimation approaches on the sizes of RCT and the target population. We aim to provide guidance for choosing appropriate average treatment effect estimators and effective variance estimation approaches under different scenarios to achieve different goals.

 \section{Methods}\label{sec:method}

\subsection{Population Average Treatment Effect}\label{sec:method_pate}
The population average treatment effect (PATE) has been defined throughout the relevant literature (e.g., \cite{brumback09, austin11,cole10, harder10, hartman2015, hirano01, imai08, kang07,kaizar2011, lunceford04, raykov12, rosenbaum87, rosenbaum83, rosenbaum84, stuart10, vansteelandt14, zhang15}). However, in the generalizability problem it is important to carefully define PATE relative to the finite or infinite size of the population of interest. We begin this section by doing so before proceeding to define several PATE estimators in Section \ref{sec:method_est}. We present methods to estimate their variance in Section \ref{sec:method_var}, including the development of a new method and detailed descriptions of two resampling-based methods. 

Based on the set up of the Rubin Causal Model, the PATE can be defined as the average difference between the counterfactual outcomes among the subjects in the target population (\cite{rosenbaum83}). We study the average treatment effect under two cases: finite and infinite population. It is necessary to differentiate finite and infinite populations, since the definition and estimation of PATE are inherently different based on the attributes of these two types of populations (\cite{kozak2005}). An example of a finite population is all the outpatient veterans with moderate or heavy alcohol intake per day in a given year, while if the population is extended across time to all possible veteran drinkers, then it could be considered infinite. We can also more realistically think about an infinite population as the distribution from which the finite population of outpatient veterans in a certain year is drawn. The observational data set in this situation can be conceived as drawing randomly from the finite population which is in turn drawn from the infinite population. 

The true target PATE is defined separately under the two types of populations. Let $Y_{1i}$ represent potential outcome variable for person $i$  if he/she had received treatment and $Y_{0i}$ represent the potential outcome variable if he/she had received control. Should the target population be a finite population, then we define the PATE to be:  
\begin{align*}
\Delta_{pop}^{*}=\mu_1^*-\mu_0^*=\frac{1}{N}\sum_{i\in \Omega_{pop}} Y_{1i}-\frac{1}{N}\sum_{i\in \Omega_{pop}}Y_{0i}
\end{align*}
where $\mu_1^*$ and $\mu_0^*$ represent the counterfactual population averages of the outcome variables, if we suppose everyone in the target population had received treatment or control respectively. $\Omega_{pop}$ is a collection of indices for the target population, and $N=|\Omega_{pop}|$ is the size of the finite population.  

Otherwise, if the target population is infinite, then we define the PATE to be: 
\begin{align*}
\Delta_{pop}=\mu_1-\mu_0=E(Y_{1i})-E(Y_{0i})
\end{align*}
where $\mu_1$ and $\mu_0$ represent the counterfactual expected values of the outcome variables of a person randomly drawn from the infinite population. 

\subsection{Estimators}\label{sec:method_est}
The point estimators for PATE that we are interested in are the same under both types of population. Because a finite population can be treated as being randomly drawn from an infinite population, an appropriate estimator for PATE under a finite population should also be appropriate for PATE under an infinite population. 

We focus on the performance of weighted estimators of PATE, particularly IPWEs, denoted $\hat{\Delta}_{IPW}$. Our IPWEs weight the trial to represent the target population via the formula suggested by Stuart $\&$ Cole (2010) \cite{cole10}, where the weights are based on data from both the target population, $\Omega_{pop}$ and the trial participants, $\Omega_{trial}$: 
\begin{align}
\label{eq:ipwe}
\hat{\Delta}_{IPW}&=
\hat{\mu}_1-\hat{\mu}_0\\
&=\left(\sum_{i\in \Omega}T_i S_i\widehat{W_i}\right)^{-1}\sum_{i\in \Omega}T_iY_iS_i\widehat{W_i}
                    &- \left(\sum_{i\in \Omega}(1-T_i)S_i\widehat{W_i}\right)^{-1}\sum_{i\in \Omega}(1-T_i)Y_iS_i\widehat{W_i}\\
\end{align}
\noindent
where $\Omega=\Omega_{trial}\cup \Omega_{pop}$ and $n=|\Omega|$ is the size of the combined trial sample and target population, $S_i$ is the indicator of being included in the trial, $T_i$ is the indicator of receiving treatment, and $\widehat{W_i}$ is the estimated weight. Roughly speaking, subject $i$ in the trial represents the number of subjects with similar covariates as subject $i$ in the target population. We postpone our discussion of how to estimate the weights until Section \ref{sec:method_ipwe} below. Notice that $S_i \neq 0$ only for subjects in the trial, so that $\hat{\Delta}_{IPW}$ is based only on outcomes observed in the trial. We calculate $\hat{\Delta}_{IPW}$ as the difference of the weighted average outcomes between the treated and control group in the trial. IPWE, $\hat{\Delta}_{IPW}$ in essence is identical to survey mean based estimator, $\hat{\Delta}_{sv}^{only}$ that is also the difference of the weighted outcomes between two groups. 

While we like the simplicity of the model-free IPW estimator, we compare it to estimators based on a linear model of the outcome: $Y_i=\eta+\bm{X_i' \beta}+T_i\gamma+\bm{\widetilde{X}_i^{'} }T_i\bm{\lambda}+\varepsilon_i$, for $i=1,...,n$ where $\eta$ is the intercept; $\bm{\beta}$ is a vector of parameters for a vector of covariates $\bm{X_i}$; $\bm{\lambda}$ is a vector of parameters for interactions between a subset of covariates $\bm{\widetilde{X}_i^{'}}$and treatment; and $\varepsilon_i$ for $i=1,...,n$ are independently identically distributed with mean $0$. 

We examine several approaches to estimating the linear model parameters. For the ordinary least squares (OLS) estimator, the parameters are estimated through ordinary least square method. The parameters in weighted OLS (WOLS) estimator are estimated with weights. The weights for subject $i$ in the trial incorporated in WOLS are the same weights, $\widehat{W_i}$ contained in IPWEs.

For survey-based estimator, the parameters are estimated through a finite population survey based method with weights (\cite{lohr2009}). The weights are again the same $\widehat{W_i}$ as in the IPWE.

Regardless of which linear model method we use, we estimate the population average treatment effect as 
\begin{align}
\label{eq:modbasedpate}
 \hat{\gamma}+\frac{1}{N}\sum_{i\in \Omega_{pop}}(\bm{\widetilde{X}_i^{'}\hat{\lambda}})
 \end{align} 
 and denoted $\hat{\Delta}_{OLS}$, $\hat{\Delta}_{WOLS}$, and $\hat{\Delta}_{modsv}$ for the OLS, WOLS, and survey methods of parameter estimation, respectively. 


\subsubsection{Construction of weights and IPWEs}\label{sec:method_ipwe}
We use PS ideas to construct weights. Rosenbaum $\&$ Rubin (1983) \cite{rosenbaum83} proposed PSs to adjust for systematic differences between treatment and control groups in observational studies, and thus ``balance" the distributions of the measured covariates between the two groups (\cite{raykov12}). We employ PSs to make the results from RCTs generalizable to a target population. In this context, we define a PS to be the probability of being selected into the trial (rather than the target population) given the measured covariates (\cite{stuart10, stuart11}). Intuitively, PSs are capable of balancing the distributions of the covariates between the trial and the target population. Thus,  they can be used to weight the trial up to some target population, such that the results from a RCT can be adjusted to be relevant to the target population. 


The weight $W_i$ is defined as, based on the weight proposed by Stuart $\&$ Cole (2010) \cite{stuart10}, \begin{align*}
W_i =\begin{cases}
 \frac{1-P(S_i=1|X_i)}{P(S_i=1|X_i)}\frac{P(S_i=1)}{1-P(S_i=1)}, & S_i=1  \\
0,& S_i=0\\
\end{cases} 
\end{align*}  
Analogous to Stuart $\&$ Cole (2010)'s \cite{stuart10} formulation, the subjects in the trial are weighted to represent the target population, while the subjects in the target population have no weight associated with them. The weights for the subjects in the trial consist of two parts. First, $\frac{1-P(S_i=1|X_i)}{P(S_i=1|X_i)}$, the ratio of the probability of being in the target population over that of being in the trial conditional on certain covariates, is aimed to make the weighted covariate distribution of the trial similar to the covariate distribution of the target population. Intuitively, we first weight the trial to reflect the entire collection of subjects, which includes the subjects both in the trial and the target population through the inverse of $P(S_i=1|X_i)$, and then weight to reflect only the target population with $1-P(S_i=1|X_i)$. Second, $\frac{P(S_i=1)}{1-P(S_i=1)}$ is the ratio of marginal probability of being selected into the trial over that of being selected into the target population. This constant ratio standardizes the weights, in a way that the weights sum up to the size of the trial in expectation. 

We estimate the two parts involved in the weights formula separately. The second part that does not include covariates can be estimated by an empirical ratio. That is, $P(S_i=1)$ is replaced by the ratio of the number of subjects in the trial to the number of subjects in the target population. The first part is a function of the conditional probability of inclusion in the trial, $P(S_i=1|X_i)$, which we call the propensity score (PS) in the generalization context. PSs can be estimated through a logistic regression model of the probability of being in the trial with the observed covariates as predictors, e.g. (\cite{stuart10}), or other nonparametric methods, e.g., (\cite{hainmueller2012}).

\subsection{Variance Estimation}\label{sec:method_var}
This section discusses the variance estimation of both model-free IPWE and estimators based on linear models. Due to the different natures of finite and infinite target populations we propose different variance estimation methods for IPWE under different types of target population . We also describe known approaches to variance estimation for model-based estimators and their relevance to the two types of the target population. 
\subsubsection{Variance Estimation of IPWE}\label{sec:method_var_ipwe}
We differentiate the variance estimators for an IPWE under different types of populations. Under the infinite population, the variance of an IPWE includes the variability of randomness of outcomes, covariates in the infinite population, and constructed weights. In contrast with the infinite population, the finite population has fixed known covariates. We do not need to account for the variability of population covariates in estimating IPWE. Thus, it is necessary for us to use different methods to estimate the variation of IPWE under different types of populations. 

The two existing methods of variance estimators for IPWE do not explicitly distinguish the types of the target population, but each is only appropriate for one estimation scenario. The estimated variance of IPWE studied in Cole and Stuart, 2010 \cite{cole10} was calculated with a finite population survey based approach, which can be obtained from standard statistical computing programs (personal communication). As for usual survey estimation practice, this variance estimation does not incorporate variance due to estimation of weights. The study in Zhang \textit{et al.}, 2015 \cite{zhang15} mentions simpler formed weighted estimators and derives the corresponding variances with asymptotic theories based on M-estimation to include the variability of the weights. Their variance estimation procedure implicitly is more suitable for the infinite target population. 

We propose variance estimators for IPWEs under both types of populations, as outlined in Table 1. We apply M-estimation to derive the variance of IPWEs under an infinite population to include all necessary variability, similar to the method presented in Zhang \textit{et al.}, 2015 \cite{zhang15}. We also use a regular bootstrap method to estimate the variance, and propose two special schemes of bootstrap to capture the variability of IPWEs with infinite target population and fixed finite target population. We describe each in detail below.

\begin{table}[H]
\caption{Variance estimation of IPWE}
\centering
\resizebox{\textwidth}{!}{
\begin{tabular}{c|ccccc|cc}
  \hline
  Variance estimation method&\multicolumn{5}{|c|}{Sources of Variance}&\multicolumn{2}{|c}{Target Populatioin}\\
& $Y$&$X$&$Weights$&$T$&$S$&Finite&Infinite\\ 
\hline
Survey-design Based& \checkmark &&&&&\checkmark &\\ 
M-estimation& \checkmark &\checkmark &\checkmark &\checkmark &\checkmark &&\checkmark\\ 
Regular Bootstrap (RB)& \checkmark &\checkmark &\checkmark &\checkmark &\checkmark &&\checkmark\\ 
Within Study Bootstrap(WSB)& \checkmark &\checkmark & \checkmark &\checkmark & &&\checkmark\\ 
Within Arm Within Study Bootstrap (WAWSB)& \checkmark &In trial only&Conditionally  &&&\checkmark &\\ 
 \hline

\end{tabular} }
\label{table:1}
\end{table}

\paragraph{Infinite Target Population}\label{sec:method_var_ipwe_inf}

The large sample variance proposed for IPWEs in Lunceford $\&$ Davidian (2004) \cite{lunceford04} inspires our derivation of the variance for our IPWEs. We apply the theory of M-estimation on our IPWE to derive the asymptotic variance (\cite{stefanski02}). The advantage of using the M-estimation method is that we can conveniently include the variability of estimating weights via logistic regression. Specifically, the probability of being selected in the trial, for subject $i$, $P(S_i=1|\bm{X_i})$ is estimated from the observed data $(S_i,\bm{X_i}), i=1,..,n$ by the logistic regression model with mean specification $e(\bm{X_i,\alpha})=\{1+exp(-\bm{X_i}^T\bm{\alpha})\}^{-1}$. It is assumed that the model for $e(\bm{X_i,\alpha})$ is correctly specified. Misspecification of the propensity score model may affect the performance of IPWEs, but is beyond the scope of this study. 

Based on M-estimation theory, then the asymptotic variance of the IPWE is $\frac{E(I_i^2)}{n}$,
where $I_i$ is defined as 
\begin{align*}\label{eq:fourth}
I_i&=\frac{1}{E\left(\frac{T_iS_i(1-e_i)}{e_i}\right)}\frac{T_iS_i(Y_i-\mu_1)(1-e_i)}{e_i}-\frac{1}{E\left(\frac{(1-T_i)S_i(1-e_i)}{e_i}\right)}\frac{(1-T_i)S_i(Y_i-\mu_0)(1-e_i)}{e_i}\\
&- \frac{1}{E\left(\frac{T_iS_i(1-e_i)}{e_i}\right)}E\left(\frac{T_iS_i(Y_i-\mu_1)(1-e_i)}{e_i}X_i^T\right)[E\left(e_i(1-e_i)X_iX_i^T\right)]^{-1}X_i(S_i-e_i)\\
&+\frac{1}{E\left(\frac{(1-T_i)S_i(1-e_i)}{e_i}\right)}E\left(\frac{(1-T_i)S_i(Y_i-\mu_0)(1-e_i)}{e_i}X_i^T\right)[E\left(e_i(1-e_i)X_iX_i^T\right)]^{-1}X_i(S_i-e_i)\tag{4}
\end{align*} 
The asymptotic variance can be estimated by ${n}^{-2}\sum_{i \in \Omega}I_i^2$. The components in $I_i$ may be estimated from the observed data, e.g., the expectations inside $I_i$ may be estimated by the simple arithmetic average across the subjects, $i=1,...,n$. 

In addition to the M-estimation method of deriving the variance, we use two bootstrap methods to estimate the variance of IPWE. We repeatedly randomly draw subjects with replacement from the combined trial and the target populations. For each bootstrap sample, IPWE is calculated. Then the variance of IPWE is obtained by taking the sample variance among the bootstrap estimators. With this method, the drawn trial sample sizes vary among the bootstrap samples. We also consider a bootstrap variation to keep the resampled trial size fixed, where for each bootstrap sample, we randomly draw trial subjects with replacement solely from the trial and draw population subjects with replacement solely from the target population. Thus, under the infinite target population, we have three methods to derive the variance of the estimators: One asymptotic method based on M-estimation and two variations of the bootstrap. 

\paragraph{Finite Target Population}\label{sec:method_var_ipwe_fin}

In addition to the standard survey sampling method, we apply a third bootstrap method to estimate the variance of IPWEs conditioning on the trial design and fixed target population. Because we consider the target population fixed and known, we do not resample the subjects in the target population. For the trial participants,  we maintain the trial design in each bootstrap sample by drawing treatment observations from the treatment group randomly with replacement and control ones from the control group in the original data set randomly with replacement. IPWEs are calculated in each bootstrap resample and the estimated variance is the sample variance across those IPWEs. 

\subsubsection{Variance Estimation of Model-Based Estimators}\label{sec:method_var_mod}
The methods of estimating the variance of model-based estimators have been established maturely. The model-based estimators are constructed as a linear combination of the parameters in \eqref{eq:modbasedpate} and the covariates in the target population. The variances of model-based estimators are computed based on the laws of variances of linear combinations, assuming covariates are fixed.  
\begin{align}
 Var\left(\hat{\gamma}+\frac{1}{N}\sum_{i \in \Omega_{pop}}(\bm{\widetilde{X}_i^{'}\hat{\lambda}})\right)&=Var\left(
 \begin{bmatrix}
 1&\frac{1}{N}\sum_{i \in \Omega_{pop}}\bm{\widetilde{X}_i^{'}}\\
 \end{bmatrix}  
\begin{bmatrix}
 \hat{\gamma}\\
\bm{\hat{\lambda}}\\
 \end{bmatrix}
 \right)\\
 &\approx \begin{bmatrix}
 1&\frac{1}{N}\sum_{i \in \Omega_{pop}}\bm{\widetilde{X}_i^{'}}\\
 \end{bmatrix}  
 Var\left(
 \begin{bmatrix}
 \hat{\gamma}\\
\bm{\hat{\lambda}}\\
 \end{bmatrix}
 \right)
 \begin{bmatrix}
 1\\
 \frac{1}{N}\sum_{i \in \Omega_{pop}}\bm{\widetilde{X}_i}\\
 \end{bmatrix}  
 \end{align}



This variance estimation method for regression-based estimators is not specifically designed for either the infinite target population or the finite target population. The set-up for defining the $Var\left(\left[
 \begin{smallmatrix}
 \hat{\gamma}\\
\bm{\hat{\lambda}}\\
 \end{smallmatrix}
 \right] \right)$ is based on an infinite target population. Thus, the procedure for calculating the variances of the estimated parameters in regression-based estimators is not inherently designed for reflecting the fixed and known population. However, step (3) is derived from (2) by assuming fixed covariates, $\bm{\widetilde{X}_i}$, $i \in \Omega_{pop}$ and the weights incorporated in the estimation of the variances are assumed to be fixed and known as well. Thus, the way we estimate the variance of the regression based estimators may underestimate the variance of the estimator if the target population is infinite. If the fixed covariate assumption is loosened and the estimation of the weights is taken into account, the variance estimator may be more suitable for an infinite target population. However, this is not standard practice for regression based estimator. 

This variance estimation method for survey-based estimators is specifically designed for the finite target population. The estimated parameters in survey based estimators, $\hat{\gamma}$, $\bm{\hat{\lambda}}$ and their variances are  based on the finite population assumption. The weights of survey based estimators contained in the estimation of $Var\left(\left[
 \begin{smallmatrix}
 \hat{\gamma}\\
\bm{\hat{\lambda}}\\
 \end{smallmatrix}
 \right] \right)$ are assumed to be fixed and known as well. Thus, this variance estimation method using (3) are designed for the finite target population. Under infinite population, the variances of model-based might be underestimated. 
 
Besides the linear combination of variances method, we apply bootstrap methods to estimate the variances of the model-based estimators as well. Following exactly the same procedures in the two bootstrap methods described in Section \ref{sec:method_var_ipwe}, these variance estimation methods similarly are suitable for the infinite target population.

\section{Simulation Study}\label{sec:sim}
We conduct simulations to compare the performance of IPWEs with that of model-based estimators. We divide our exploration into two simulation studies: one is to investigate the effect of the selection effect and heterogeneity of treatment effect in a large trial sample and the other one is to examine the effect of sample size and proportion. Both simulations have the same set up and only differ in choices of parameter settings. 
\subsection{Simulation Set-up}\label{sec:sim_setup}
We set up the simulations to include a binary treatment variable ($T_i=0$ or $1$), one continuous covariate ($X_i$) and a continuous response ($Y_i$), for simplicity. A study indicator ($S_i$) mimics a realistic scenario by dividing the data into two parts --- the trial data ($S_i=1$) and the target population data ($S_i=0$). 

\begin{figure}[H]
\centering
\includegraphics[width=0.7\textwidth]{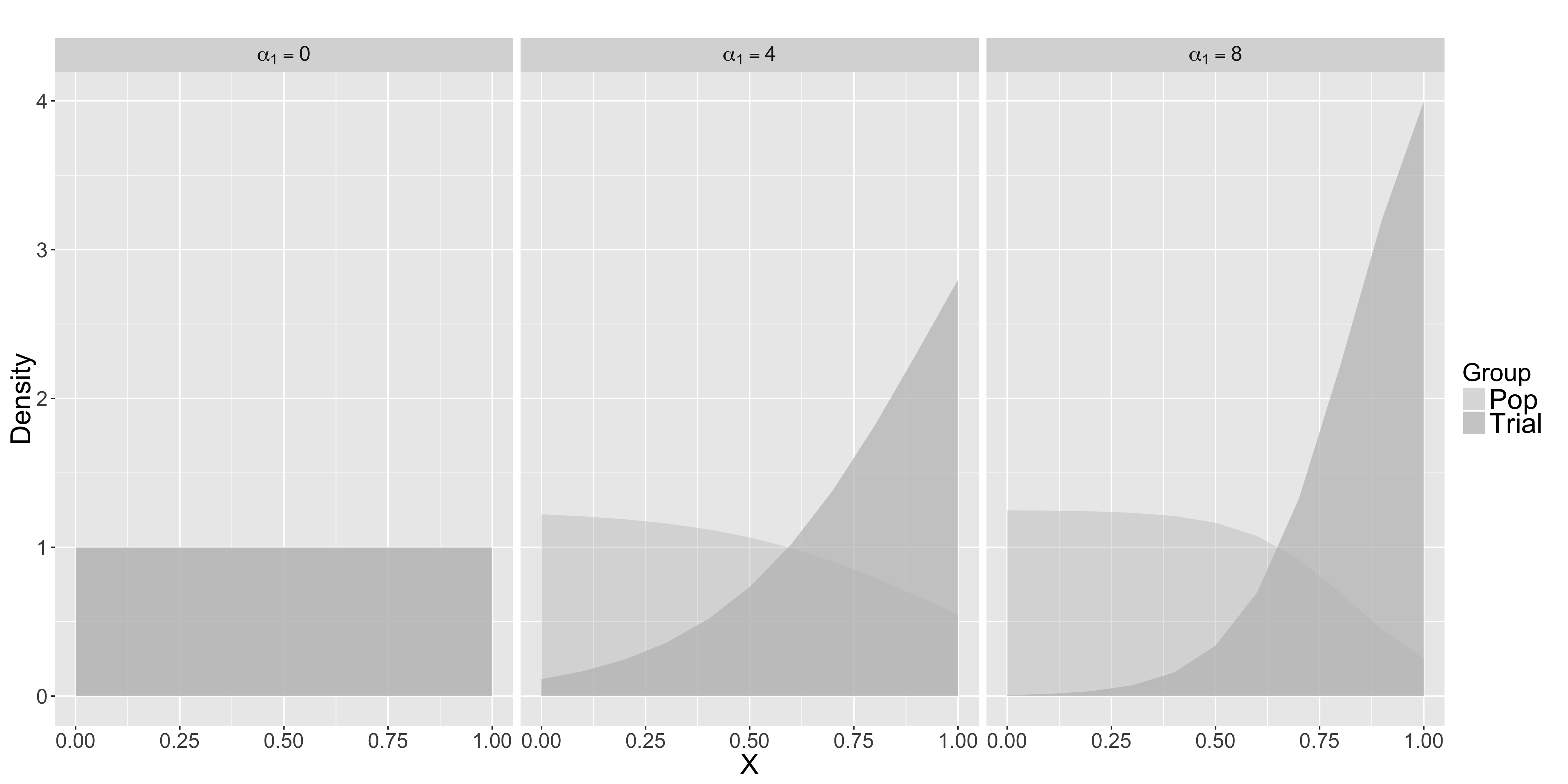}
\caption{The conditional probability densities for $X$ in the trial (``Trial", dark grey) and the target population (``Pop", light grey) for three possible values of $\alpha_1$ in Equation (\ref{eq:1}).}
\label{figure:00}
\end{figure}

The continuous covariate $X_i$ is generated as a Uniform random variable between 0 and 1 for each subject $i \in \Omega$. The trial assignment $S_i$ is generated according to a Bernoulli distribution with the true propensity score, $P(S_i=1|X_i)$,  modeled as in a logistic regression on $X_i$:
\begin{equation}
P(S_i=1|X_i)=\{1+\text{exp}(-\alpha_0-\alpha_1\X)\}^{-1}
\label{eq:1}
\end{equation}
The logistic model is designed to mimic a potential treatment moderator that may be distributed differently in the trial and the target population. In general, values of $\alpha_1$ close to zero produce considerable overlapping of the conditional distributions of $X$ across the trial and the target population. As $\alpha_1$ grows in magnitude, the overlapping amount decreases. Figure \ref{figure:00} depicts this trend. When $\alpha_1=0$ (left panel), there is no selection and the conditional distribution $X|S=s$ is identical across the two groups. As $\alpha_1$ increases to $8$ (right panel), the two conditional distributions differ noticeably, with most subjects in the trial and few subjects in the target population having large values of $X$. 
\noindent
For only the subjects $i$ selected in the trial, i.e., $S_i=1$, the active treatment exposure $T_i=1$ is randomly assigned to exactly half of those units in the trial; the other half are assigned to the control treatment $T_i=0$. For only the subjects in the trial as well, the response variable $Y_i$ is generated according to 
\begin{equation}
Y_i=\beta_0+\beta_1 X_i+\beta_2 T_i+\beta_3 X_iT_i+\varepsilon_i  \text{,} \quad 
\varepsilon_i \stackrel[]{ind}{\sim} N(0,\sigma^2)
\label{eq:2}
\end{equation}
Our simulated data thus mimics the real world situation where individuals in the target population may not have been treated and do not have outcome data available. 

We have two simulation constructions to distinguish the types of the target population: finite and infinite. We employ a single layer simulation set up to mimic the reality that the target population is an infinite population. We use a double-layer simulation to mimic the reality that the target population is a specific finite population.

\subsubsection{Infinite Target Population - Single-layer Simulation}\label{sec:sim_setup_inf}
Within the single layer simulation, each Monte Carlo (MC) data set represents subjects randomly drawn from an infinite population whose covariate distribution follows a Uniform distribution. 3000 MC data sets are generated and . In each of the MC data sets, covariate $X$, trial assignment $S$, treatment assignment $T$, and the outcome $Y$ are generated, such that the trial and the population subjects vary across all 3000 MC data sets. 
 
\subsubsection{Finite Target Population - Double-layer Simulation}\label{sec:sim_setup_fin}
The double-layer simulation is built on the single-layer simulation. Since under the finite target population, the target population subjects are fixed, we add a nested layer within each single-layer dataset that shares the same finite target population subjects, but generate 10 MC replications of different trial subjects. For each of the 10 trial replications in the nested layer, we generate covariate $X$ based on the conditional distribution of the covariate, $P(X|S=1)$ and then generate the treatment $T$ and outcome $Y$ as usual according to Equation \ref{eq:2}. In other words, the outside layer of the simulation contains 3000 distinct target populations, while in the nested inside layer, there are 10 replications sharing same target population, but varying covariates and outcomes in the trial for a total of 30000 simulated RCTs. The replications within the nested layer represent simulated replicates of a RCT, each of which is generalized to a fixed finite target population. This mimics the real-life situation where many trials are possible, but only one finite target population is of interest. 

\subsection{Simulation Estimators}\label{sec:sim_est}
Three major types of PATEs are computed in the simulations.  Regression model-based estimators are the ordinary least squares estimator, $\widehat{\Delta}_{OLS}$ and the linear model estimator including the propensity adjustment, $\widehat{\Delta}_{WOLS}$ in \ref{eq:modbasedpate} of Section \ref{sec:method_est}. The survey model-based estimator is $\widehat{\Delta}_{modsv}$. The IPWEs are $\widehat{\Delta}_{IPW}$ and $\widehat{\Delta}_{sv}^{only}$ in Equation \ref{eq:ipwe} of Section \ref{sec:method_est}. The model-based estimators not only rely on the specification of PS model, but also that of the outcome analysis model. Thus, for each type, we include estimators under the case of both correctly and incorrectly specified analysis models. The superscript, $cor$, indicate estimators computed under the correct outcome analysis model (i.e., the model that was used to generate the data). The misspecified analysis models have no superscript and are calculated  without the interaction term between treatment indicator and covariate (i.e., $\hat{\lambda}=0$ in Equation \ref{eq:modbasedpate}), i.e., assuming no treatment heterogeneity. 

As discussed in the variance estimation section \ref{sec:method_var}, in the single layer simulation, three types of variance estimators are calculated for each PATE estimator: the regular bootstrap, the special bootstrap with fixed $S$, and a  linear combination of variances for model based estimators or M-estimation for IPWEs. In the double layer simulation, we apply the special bootstrap with fixed $S$ and $T$ to estimate the variances for IPWEs and we use a linear combination of variances for the rest of the model-based treatment effect estimators as we do in the single layer simulation set up. 

\subsection{Parameters for Simulation Study 1: study of effect of Selection effect and heterogeneity of treatment effect }\label{sec:sim_param1}
 We examine the performance of the treatment effect and variance estimators for different amounts of covariate overlap and treatment effect heterogeneity. As described above, $\alpha_1$ in Equation \ref{eq:1} controls the degree of overlap between the covariate distribution of the trial and the target population or the size of the selection effect. Thus, we vary the values of $\alpha_1$ to examine the performance of the estimators as the overlap changes. Table \ref{tb:1} summarizes the 3 settings of $\alpha s$ we used and the resulting simulated data properties. In each case, we determine $\alpha_0$ to achieve a pre-specified probability for subjects to be selected in the trial. $\Delta_p$, calculated as the difference of the average of propensity scores between the trial and the target population, measures the overlapping between the covariate distribution of the target populations and the trial, as suggested by Stuart, \textit{et al.}, 2011\cite{stuart11}.  

\begin{table}[H]
\centering
\begin{tabular}{c|cccc}
  \hline
Type & $\alpha_0$&$\alpha_1$&$P(S=1)$&$\Delta_p$\\ 
 \hline
 1&-1.39&0&0.2&0\\
 2&-3.76&4&0.2&0.15\\
 3&-6.62&8&0.2&0.38\\
 \hline
\end{tabular} 
\caption{The settings of $\alpha s$ and resulting $\Delta_p$, the difference of the average of propensity scores between the trial and the target population.}
\label{tb:1}
\end{table}

$\beta_3$ in the outcome analysis model (Equation \ref{eq:2}) represents the heterogeneous effect of treatment. We also vary the value of $\beta_3$ to examine the performance of the estimators for different magnitudes of heterogeneity. The true PATE, $\beta_2+\beta_3E(X|S=0)$, is preassigned, and $\beta_2$ is determined so as to achieve this PATE for each setting for $\beta_3$. $\beta_0$ is set to be 0, since it only serves as a constant. Since $\beta_1$ is only prognostic, it should not affect the performance of the treatment effect estimators and variance estimators, $\beta_1$ is constantly set to be 0.3. We select the values of  $\beta_1=0.3$, $\beta_2$, and $\beta_3=[-1,1]$ to be similar to the outcome analysis models by scaling the estimated parameter values of the outcome analysis model in the empirical example in Section 4. Table \ref{tb:2} summarizes the combinations of $\alpha s$ and $\beta s$ we used. 
\begin{table}[H]
\centering
\begin{tabular}{cc|ccc|c}
  \hline
  Setting &&\multicolumn{3}{|c|}{$\beta_2$}&\\

 & $\beta_1$&$\alpha_1=0$&$\alpha_1=4$&$\alpha_1=8$&$\beta_3$\\ 
 \hline
 1&0.3&0.2&0.14&0.12&-1\\
 2&0.3&0.1&0.06&0.04&-0.8\\
 3&0.3&0&-0.03&-0.05&-0.6\\
 4&0.3&-0.1&-0.12&-0.13&-0.4\\
 5&0.3&-0.2&-0.21&-0.22&-0.2\\
 6&0.3&-0.3&-0.3&-0.3&0\\
 7&0.3&-0.4&-0.39&-0.38&0.2\\
 8&0.3&-0.5&-0.48&-0.47&0.4\\
 9&0.3&-0.6&-0.57&-0.55&0.6\\
 10&0.3&-0.7&-0.66&-0.64&0.8\\
 11&0.3&-0.8&-0.74&-0.72&1\\
 12&0.3&-100.3&-89.38&-84.42&200\\
 \hline
\end{tabular} 
\caption{The settings of $\beta s$ across different settings of $\alpha_1$. In each case, values of $\alpha_0$ are set as in Table \ref{tb:1}, and the combined size is $n_{total}=3000$.}
\label{tb:2}
\end{table} 
 
\subsection{Parameters for Simulation Study 2: study of effect of sample and study size} \label{sec:sim_param2}
We examine how the performance of the treatment effect and variance estimators changes with RCT and target population size. We vary the total combined size of the trial and the target population, $n_{total}$ and the expected sample proportion, $P(S=1)$, to control the target population size and the trial size. We incrementally increase the sample proportion and meanwhile decrease the combined size, $n_{total}$, while we keep $\Delta_p$ constant. We set $\Delta_p$ as 0.15, which is its value of $\alpha_1=4$ and $P(S=1)=0.2$. We keep the value of the true PATE the same as Simulation Study 1. Note that the $\alpha s$ depend on $\Delta_p$ and the sample proportion, while the $\beta s$ depend on $\alpha s$ and the true PATE. Table \ref{tb:5} summarizes the settings of $\alpha s$ and $\beta s$ we used and the resulting simulated data properties.

\begin{table}[H]
\centering
\begin{tabular}{c|cc|ccc|cc|cc}
  \hline
 Setting& $\alpha_0$&$\alpha_1$&$\beta_1$&$\beta_2$&$\beta_3$&$n_{total}$&\begin{tabular}{@{}c@{}}Sample \\ Proportion\end{tabular} &\begin{tabular}{@{}c@{}}Treatment \\ Effect\end{tabular}&$\Delta_p$ \\ 
 \hline
 1&-3.76&4&0.3&0.14&-1&3000&20 $\%$&-0.3&0.15\\
 \hline
 13&-3.76&4&0.3&0.14&-1&1000&20$\%$&-0.3&0.15\\
 14&-3.76&4&0.3&0.14&-1&500&20$\%$&-0.3&0.15\\
 15&-3.76&4&0.3&0.14&-1&200&20$\%$&-0.3&0.15\\
 16&-1.5&3&0.3&0.09&-1&500&50$\%$&-0.3&0.15\\
 17&0.01&4.5&0.3&-0.05&-1&500&85$\%$&-0.3&0.15\\
  18&0.56&9&0.3&-0.18&-1&500&95$\%$&-0.3&0.15\\

 \hline
\end{tabular} 
\caption{The settings of the parameters for the study of the effect of sample size and proportions. }
\label{tb:5}
\end{table}
\subsection{Measures of Performance}\label{sec:sim_measure}
We compute several measures to describe and evaluate the properties of the PATE estimators for both single and double-layer simulations. For the single-layer simulations, bias is estimated by averaging the differences between the treatment effect estimators and the true population average treatment effect, $\beta_2+\beta_3E(X|S=0)$. Average standard error (Ave SE) is computed as the average of the estimated standard deviations of the PATE estimators described in Section \ref{sec:method_var} across 3000 MC data sets. Monte Carlo standard deviation (MC SD) is computed as the standard deviation of the treatment effect estimators across the 3000 MC data sets using the usual sample standard deviation formula. MC SD is a numeric estimate of the true standard deviation of the estimators, and thus it is treated as the measure of the true standard deviation of the estimators. Finally, two types of 95$\%$ confidence interval coverage rates are computed across the 3000 MC data sets. We calculate a separate confidence interval for each type of target population: infinite and finite target population. The infinite population confidence interval is designed for covering the true PATE, $\beta_2+\beta_3E(X|S=0)$, while the finite population confidence interval is for covering each particular simulation PATE, $\beta_2+\beta_3\bar{X}_{pop}$. 

Similar measures are computed for double-layer simulations. For the double-layer simulations, bias is estimated by averaging the differences instead between the treatment effect estimators and each specific simulation  PATE, $\beta_2+\beta_3\bar{X}_{pop}$ across a total of 30000 MC data sets. Note that distinct MC data sets have distinct $\bar{X}_{pop}$. Ave SE and two types of confidence coverage follow the same computing procedures as in the single-layer simulations, but instead they are the averages across the 30000 MC data sets. MC SD is the average of the 3000 standard deviation, each of which is the standard deviation of the treatment effect estimators across the 10 replications sharing same target population.

\subsection{Simulation Results}\label{sim_result}

For every combination of $\alpha$ and $\beta$, the estimated results and the related measures of both the model-based estimators and IPWE are computed under the single-layer simulation construction designed to assess properties for the infinite target population and also double-layer simulation construction correspondingly designed for the finite target population. We first present results for both constructions for one exploration of the effect of selection and heterogeneity, Simulation 1, before turing to our Simulation 2 study of the effect of sample size.  

\subsubsection{Study of Effect of Selection and Heterogeneity of Treatment Effect}\label{sim_result_1}
\paragraph{Single-layer Simulation Results}\label{sim_result_1_inf}

\begin{figure}[htp]
\centering
\includegraphics[height=4.5cm]{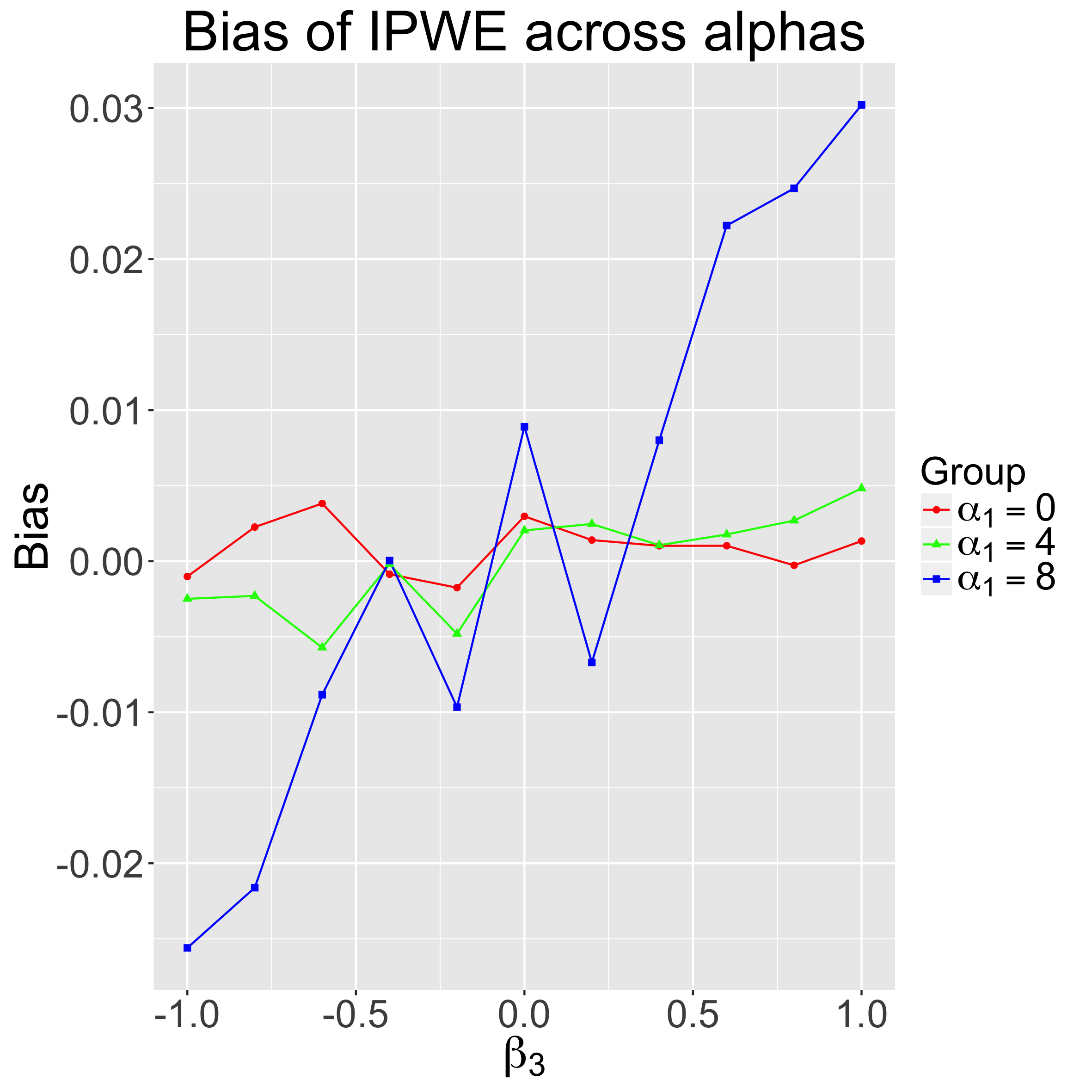}\quad
\includegraphics[ height=4.5cm]{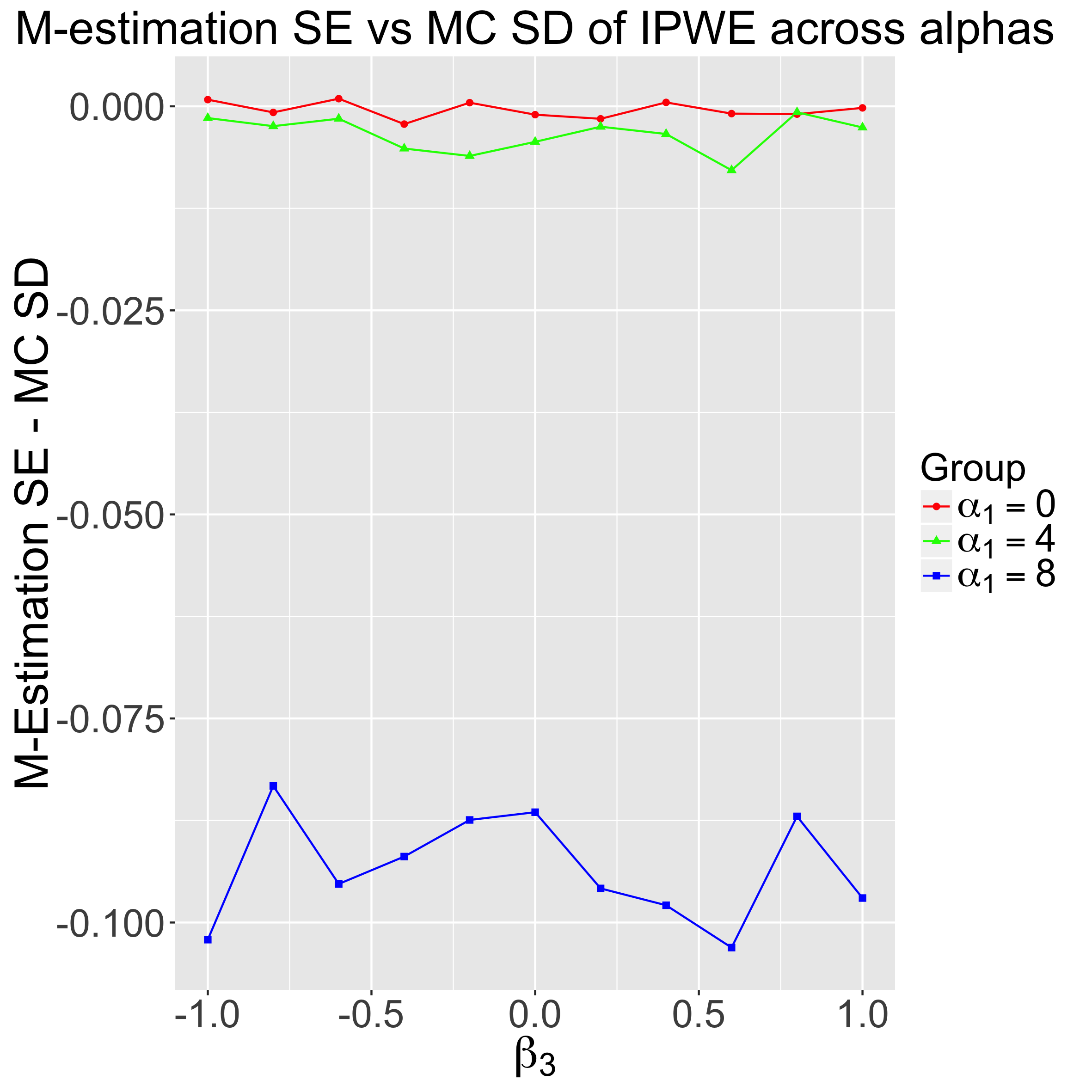}
\medskip

\includegraphics[height=4.5cm]{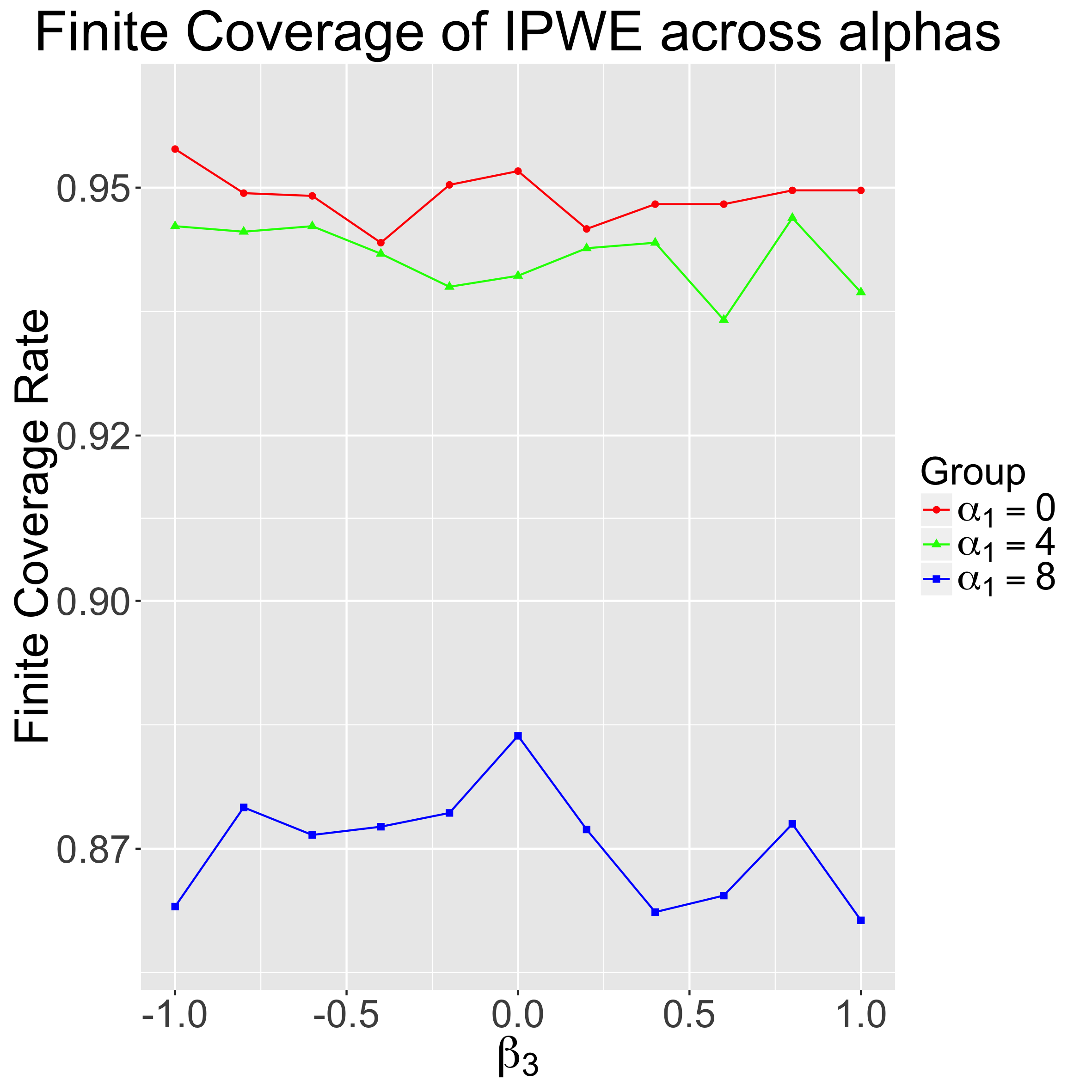}\quad
\includegraphics[ height=4.5cm]{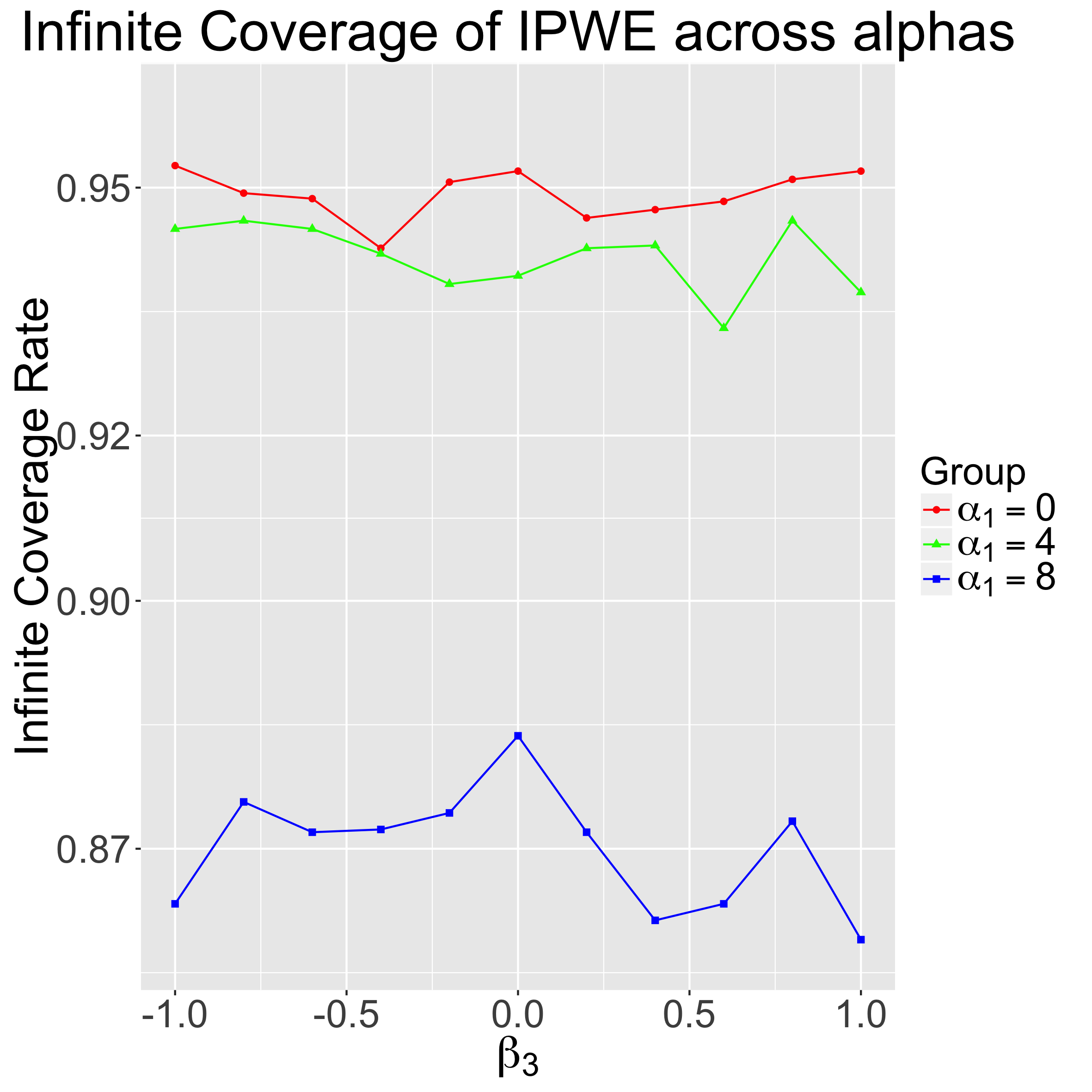}
\caption{The first row of the panel is bias and the differences between Ave SE and MC SD across $\beta_3$ values and $\alpha_1=0,4,$ and $8$; the second row of the panel is two types of 95$\%$ coverage rates of IPWE across $\beta_3$ values and $\alpha_1=0,4,$ and $8$.}
\label{figure:2}
\end{figure}

Under the single-layer simulation construction, we first look at one generic situation which has moderate effect of selection and heterogeneity with$\alpha$ setting 2, where $\alpha_1=4$ in Table \ref{tb:1} and $\beta$ setting 3, where $\beta_3=-0.6$ in Table \ref{tb:2}. Table \ref{tb:3} summarizes the results of the corresponding simulation. Then we compare the trend of the performances of the PATE estimators as we change the effect of selection and heterogeneity. Figures \ref{figure:2}  summarize how the results change with $\alpha_1$ and $\beta_3$. 

\begin{table}[H]
\centering
\resizebox{\textwidth}{!}{
\begin{tabular}{c||cc|ccc|ccc|ccc}
  \hline
  &&&\multicolumn{3}{|c|}{\textbf{Parametric Variance Estimator}}&\multicolumn{3}{|c|}{\textbf{Fixed S Bootstrap}}&\multicolumn{3}{|c}{\textbf{Bootstrap}}\\
 Est& Bias & MC SD & Ave SE & \begin{tabular}{@{}c@{}}Finite \\ Coverage\end{tabular} & \begin{tabular}{@{}c@{}}Infinite \\ Coverage\end{tabular} & Ave SE & \begin{tabular}{@{}c@{}}Finite \\ Coverage\end{tabular} & \begin{tabular}{@{}c@{}}Infinite \\ Coverage\end{tabular} & Ave SE & \begin{tabular}{@{}c@{}}Finite \\ Coverage\end{tabular} & \begin{tabular}{@{}c@{}}Infinite \\ Coverage\end{tabular} \\ 
  \hline
$\widehat{\Delta}_{OLS}$ & -0.166 & 0.082 & 0.082 & 0.48 & 0.47 & 0.082 & 0.47 & 0.47 & 0.082 & 0.47 & 0.47 \\ 
  $\widehat{\Delta}_{OLS}^{cor}$ & -0.005 & 0.129 & 0.130 & 0.95 & 0.95 & 0.129 & 0.95 & 0.95 & 0.129 & 0.95 & 0.95 \\ 
  $\widehat{\Delta}_{WOLS}$ & -0.007 & 0.141 & 0.082 & 0.74 & 0.74 & 0.138 & 0.94 & 0.94 & 0.138 & 0.94 & 0.94 \\ 
  $\widehat{\Delta}_{WOLS}^{cor}$ & -0.005 & 0.142 & 0.082 & 0.74 & 0.74 & 0.140 & 0.94 & 0.94 & 0.140 & 0.94 & 0.94 \\ 
  $\widehat{\Delta}_{modsv}$ & -0.007 & 0.141 & 0.139 & 0.95 & 0.94 & 0.138 & 0.94 & 0.94 & 0.138 & 0.94 & 0.94 \\ 
  $\widehat{\Delta}_{modsv}^{cor}$ & -0.005 & 0.142 & 0.139 & 0.94 & 0.94 & 0.140 & 0.94 & 0.94 & 0.140 & 0.94 & 0.94 \\ 
  $\widehat{\Delta}_{sv}^{only}$ & -0.006 & 0.141 & 0.140 & 0.95 & 0.95 & 0.139 & 0.94 & 0.94 & 0.139 & 0.94 & 0.94 \\ 
  $\widehat{\Delta}_{IPW}^T$ & -0.006 & 0.141 & 0.140 & 0.95 & 0.94 & 0.140 & 0.94 & 0.94 & 0.140 & 0.94 & 0.94 \\ 
   \hline
\end{tabular}}
\caption{Simulation results for $(\alpha_0,\alpha_1)=(-3.76,4)$ and $\beta_3=-0.6$ under single-layer simulation set-up. See Section 3.2 for a description of the estimators and Section 3.1 for a description of the summaries.}
\label{tb:3}
\end{table}

Table \ref{tb:3} clearly shows the difference of bias across all the estimators. Without knowing the correctly specified analysis model, the unweighted regression model-based estimators $\widehat{\Delta}_{OLS}$ has the largest bias among all the treatment effect estimators. The model-based estimators with correct analysis model have almost the same amount of bias and have less bias than the model-based estimators with incorrect analysis model do, which have consistent amount of bias also. The bias of the IPWEs is between the model-based estimators with correctly specified analysis model and those with incorrectly specified model. 

MC SD column reflects the true variability of each PATE estimators. $\widehat{\Delta}_{OLS}$ has the smallest MC SD, but with the largest bias, we  practically do not prefer it. Thus, we do not discuss it regarding the other property comparison. With correct analysis model,  $\widehat{\Delta}_{OLS}^{cor}$ has the smallest standard deviation. The rest of the model-based PATE estimators and IPWEs have similar MC SD.  

Table \ref{tb:3} compares the performances of the variance estimation methods: Parametric M-estimation, survey-based and model-based linear combination estimators, Fixed S bootstrap, and regular bootstrap methods as described in Section \ref{sec:method_var}. In this single-layer simulation construction, the standard deviation estimate of IPWEs from the Parametric methods and the two bootstrap methods are all slightly smaller than, but very close to the ``true" standard deviation, MC SD, except for the WOLS estimators, $\widehat{\Delta}_{WOLS}$ and $\widehat{\Delta}_{WOLS}^{cor}$. In this case, the linear combination of variances method severely under-estimates the true standard deviation. This is not entirely surprising, since the weights in a weighted regression have a variance interpretation quiet different from population representation.

The finite and infinite population coverages are almost the same for each variance estimation method across the treatment effect estimators. There are two types of estimators for which the coverage is poor, WOLS under correct or incorrect analysis model with Parametric variance methods. With under-estimates of the variances, the corresponding coverages are below the nominal level. All the other intervals have coverages that achieve or nearly achieve $95 \%$. 

The same comparative performance hold across all the $\alpha_s$ and $\beta_s$ settings. 

Because it has good properties and is robust to model mis-specification, we focus on the IPWEs. Both heterogeneous effect of treatment (as determined by $\beta_3$) and the similarity between the trial and the target population covariate distributions (as determined by $\alpha_1$) determine the scale of  the bias of the IPWE, as shown in the left panel of Figure \ref{figure:2}. If there is a considerable amount of overlap between the covariate distributions of the trial and the target population (i.e., $\alpha_1=0$), the trial can fully represent the target population through weighting by IPWE, and thus the bias of the IPWE is pretty small. Furthermore, if  $\beta_3=0$, there will be no heterogeneity and thus no bias, regardless of the values of the other parameters. As $\alpha_1$ increases or the absolute value of $\beta_3$ increases, the absolute value of the bias of the IPWE increases. (Note that the deviations from smooth lines in this figure are due to MC variation.) We expected to see these trends. If there is not enough overlap between the covariate distributions (as for large $\alpha_1$), some part of the target population is unable to be well-represented by the trial through weighting and thus the IPWE is biased. If paired with large heterogeneous treatment effect, then the bias may be worse. We can clearly see that the absolute value of the bias of IPWE climbs up as $\beta_3$ increases. This pattern is clearest for $\alpha_1=8$, but is also present for smaller $\alpha_1=4$. 

\begin{figure}[H]
\centering
\includegraphics[width=0.5\textwidth]{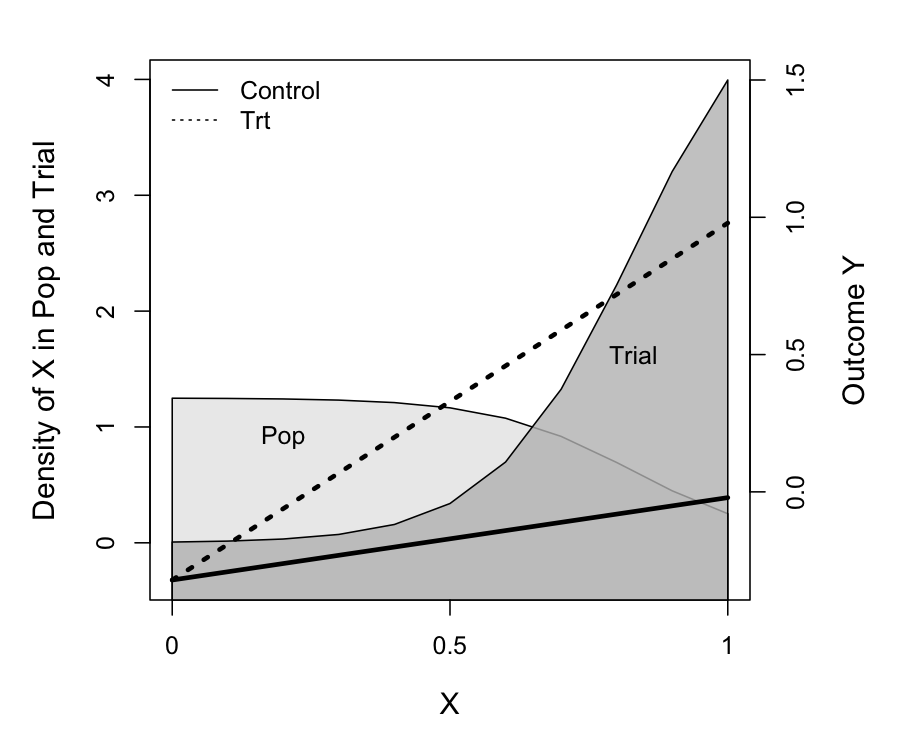}
\caption{Plot with $(\alpha_0,\alpha_1)=(-6.62,8)$ and positive heterogeneous effect. The trial subjects most likely are assigned with large scale of covariates, while the target population subjects most likely have small covariates. With positive heterogeneous effect, subjects with large covariates tend to have large treatment effect. }
\label{figure:3}
\end{figure}
The heterogeneous effect of treatment and the similarity between the trial and the target population also indicate the sign of the bias of the IPWE. As shown in the left panel of Figure \ref{figure:2}, with positive and large heterogeneous treatment effects, the subjects who have large covariates tend to have large treatment effects. Figure \ref{figure:3} depicts these phenomena for simulation setting 11 with $\beta_3=1$ and $\alpha_1=8$. Because $\alpha_1$ is positive, most of the mass for the density of $X$ in the trial is for greater values than most of the mass for the target population. Because $\beta_3$ is positive, the treatment effect (the difference between the solid and dotted lines) is also larger for these large values of covariate $X$. As $\alpha_1$ increases, part of the target population, which are assigned with smaller covariates may not be represented by the trial. In other words, the weighted trial only represents most of the subjects in the target population with large covariates. Thus, as positive heterogeneous treatment effect $\beta_3$ increases, the bias increases in the positive direction. Vice versa, a negative heterogeneous treatment (negative $\beta_3$) equally means that the subjects with large covariates may have small treatment effects. Then the average population treatment effect is estimated among the population who intend to have large covariates. Thus, the  treatment effect estimated by the IPWE might be smaller than the true PATE, or negatively biased.

The trend of various variance estimation methods depend on the parameter values. The performance of the M-estimation variance method is related to the value of $\alpha_1$, which controls the similarity of the covariate distributions of the target population and the trial. With closer to 0 $\alpha_1$, there is a considerable amount of overlapping between the target population and the trial and there is almost no difference between M-estimation estimates of IPWE and its corresponding MC SD as shown by the red line in the top right figure of Figure \ref{figure:2}. That is, the variance estimate is approximately unbiased. As $\alpha_1$ increases, the bias increases as well and the M-estimation method underestimates the ``true" variability of IPWE, which is shown most clearly by the blue line in the top right figure of Figure \ref{figure:2}. The heterogeneous effect, $\beta_3$ does not affect the accuracy of the M-estimation variance estimation method. 

The two bootstrap methods, the regular one and the special one with fixed the trial indicator $S$ have similar performance. Across all the scenarios, the estimated standard deviations from both bootstrap methods are almost identical and they are very similar with the parametric estimation results, except for the estimates for the weighted regression PATEs. In that case, the bootstrap methods perform better than the Parametric method. 

The trends in bias and SE estimator accuracy imply the trends we see in CI coverage. Under the situation that the target population and the trial is similar enough (i.e. $\alpha_1$ close to 0 as in the highest two lines in Figure \ref{figure:2}), the 95$\%$ confidence interval for the IPWE captures both the infinite population average treatment and the specific target population average treatment nearly at the nominal level. Since there is little difference in the performance of the variance estimators, there is also almost no difference among the coverage rates for all three variance estimation method for the IPWE. As $\alpha_1$ increases, the coverage rates apparently move below 95$\%$ as shown by the differences between the three lines in the center and right panels of Figure \ref{figure:2}. This is not surprising since the bias increases and the variance estimation methods perform poorly as $\alpha_1$ increases. 

\paragraph{Double-layer Simulation Results}\label{sim_result_1_fin}

\begin{figure}[H]
\includegraphics[width=0.3\textwidth]{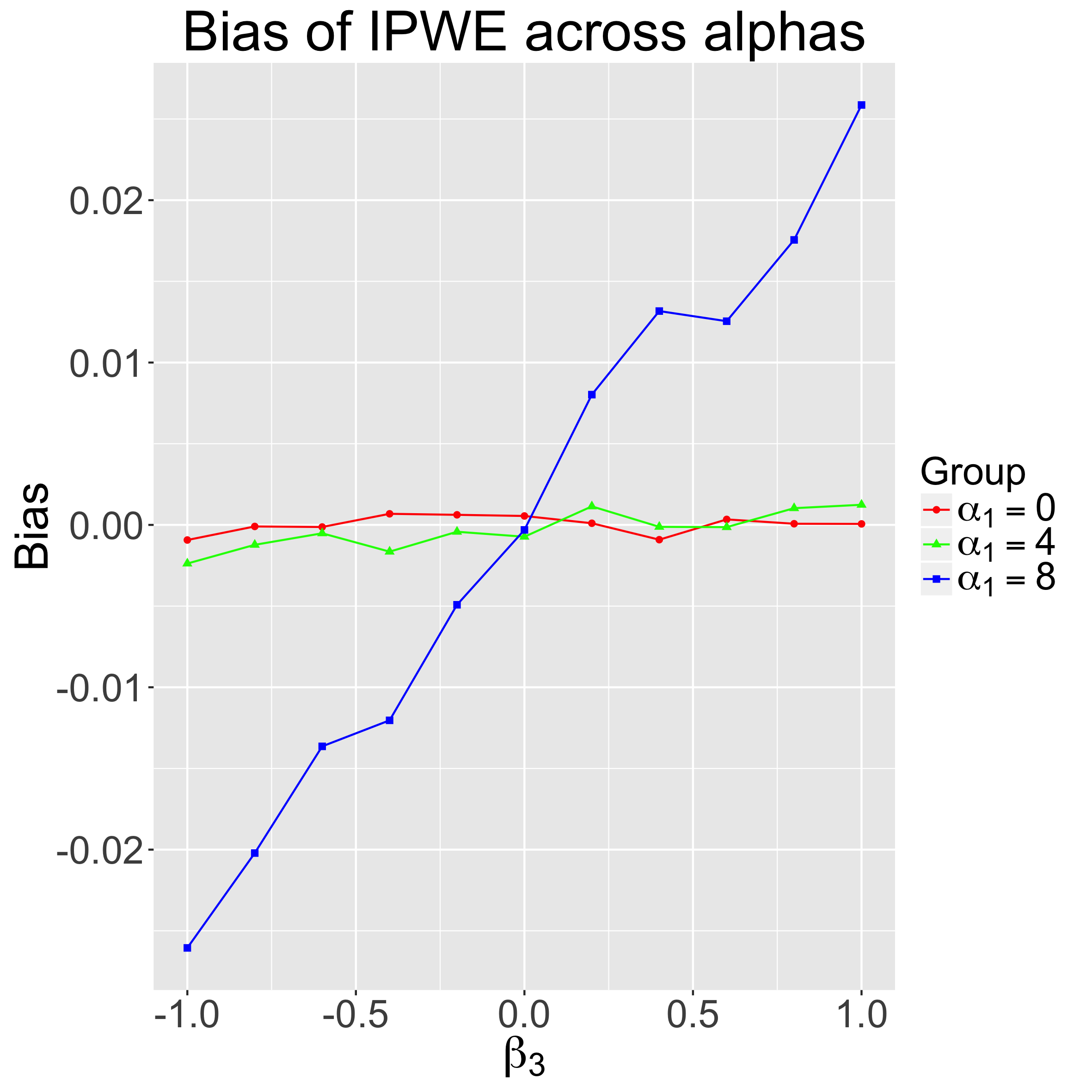}
\includegraphics[width=0.3\textwidth]{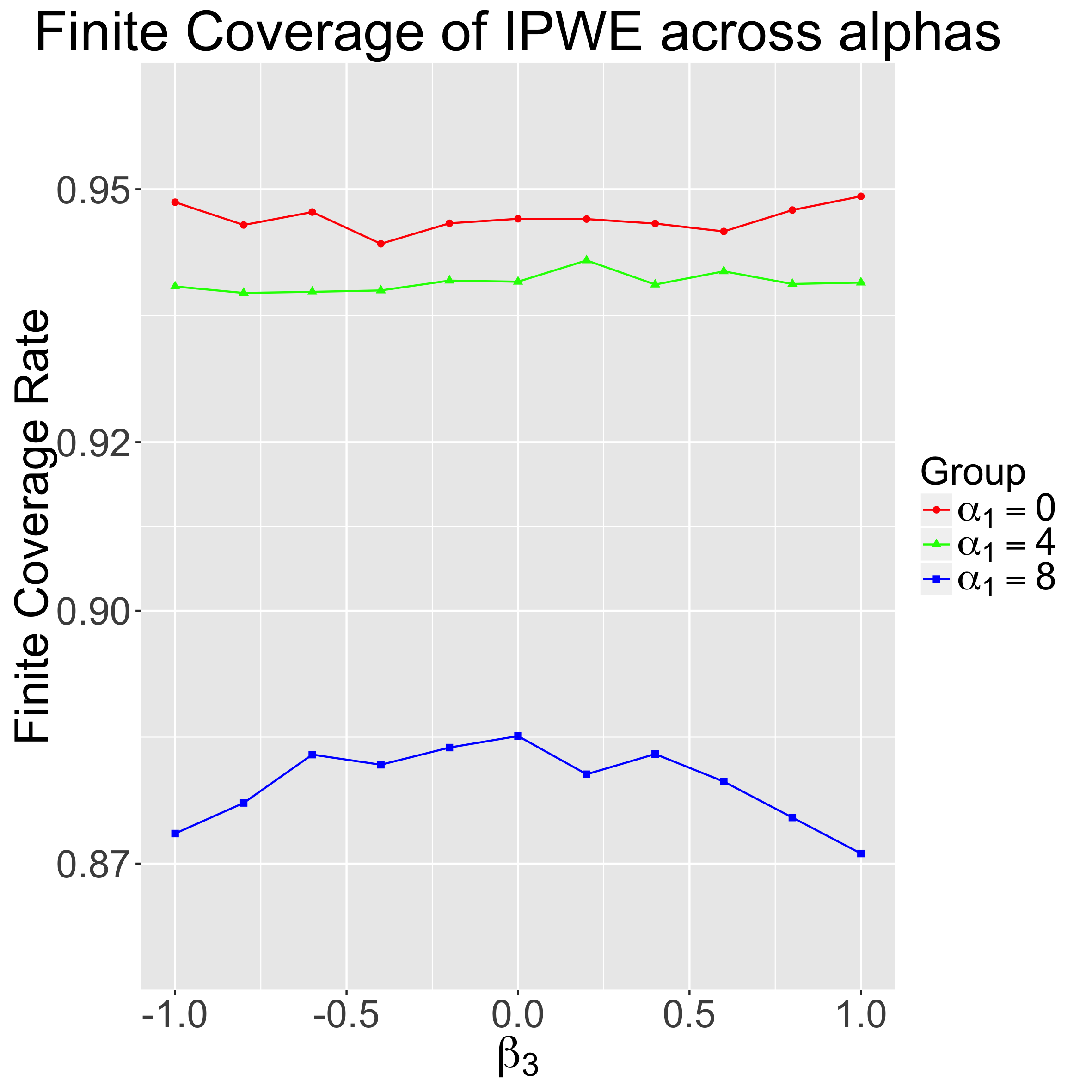}
\includegraphics[width=0.3\textwidth]{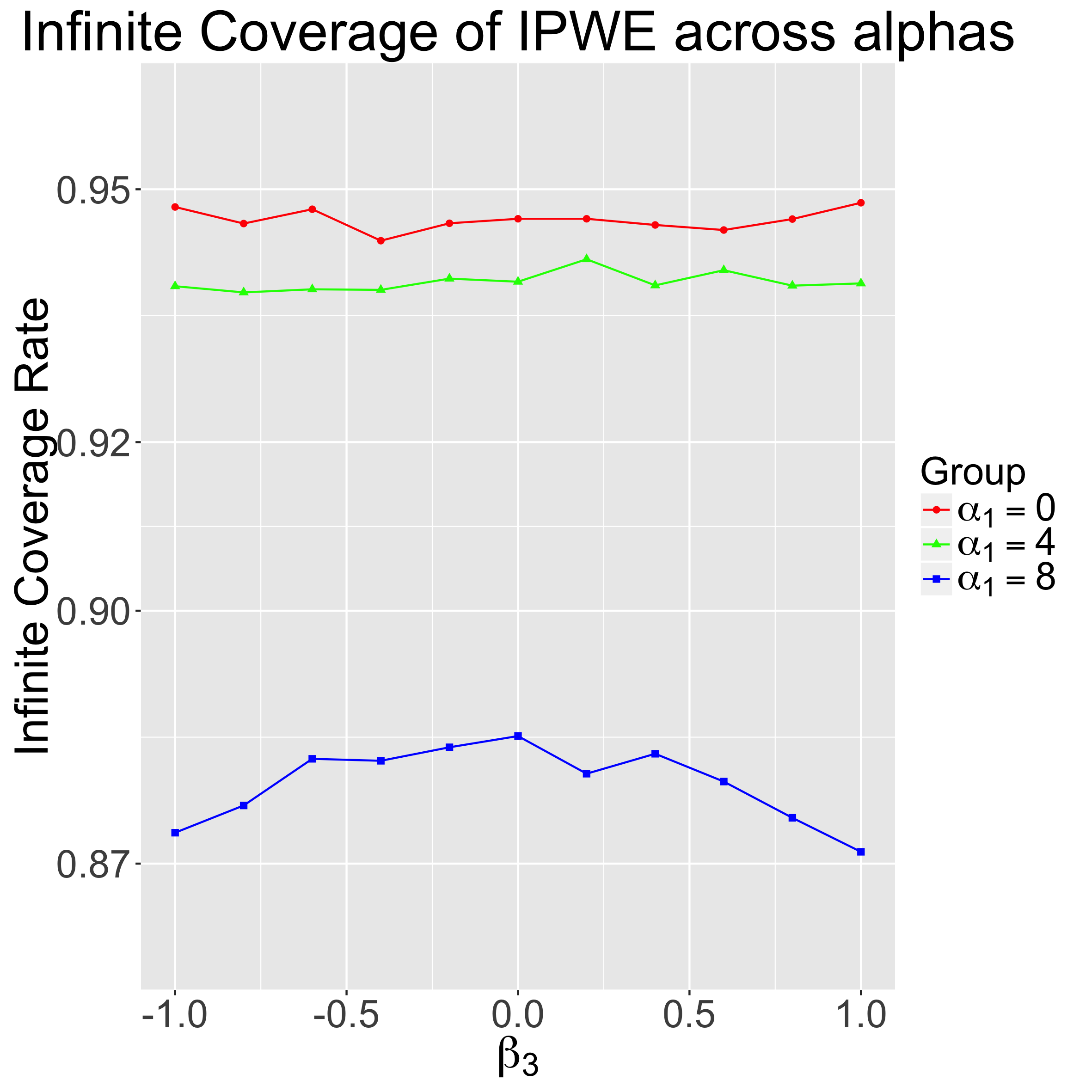}
\caption{Bias and two types of 95$\%$ coverage rates of IPWE across $\beta_3$ values and $\alpha_1=0,4,$ and $8$. }
\label{figure:4}
\end{figure}

Using double-layer simulation set-up, similar as single-layer simulation analysis, we first look at one generic situation which has moderate effect of selection and heterogeneity with$\alpha$ setting 2, where $\alpha_1=4$ in Table \ref{tb:1} and $\beta$ setting 3, where $\beta_3=-0.6$ in Table \ref{tb:2}. Table \ref{tb:4} summarizes the results of the corresponding simulation. Then similar to the comparison of results across all scenarios in the single-layer construction, we examine the performances of the treatment effect estimators in the double-layer construction. Both the magnitudes and the trends we showed regarding the bias in the singl-layer construction are maintained in the results of the double-layer coustruction. We compare the trend of the performances of the PATE estimators as we change the effect of selection and heterogeneity. Figures \ref{figure:4} summarizes how the results change with $\alpha_1$.

\begin{table}[H]
\centering
\begin{tabular}{c|rr|rrr}
  \hline
 Est& Bias & MC SD & Ave SE & \begin{tabular}{@{}c@{}}Finite \\ Coverage\end{tabular}  & \begin{tabular}{@{}c@{}}Infinite \\ Coverage\end{tabular}  \\ 
  \hline
$\widehat{\Delta}_{OLS}$ & -0.166 & 0.080 & 0.082 & 0.48 & 0.48 \\ 
  $\widehat{\Delta}_{OLS}^{cor}$ & 0.001 & 0.127 & 0.130 & 0.95 & 0.95 \\ 
  $\widehat{\Delta}_{WOLS}$ & -0.002 & 0.139 & 0.082 & 0.74 & 0.74 \\ 
  $\widehat{\Delta}_{WOLS}^{cor}$ & 0.001 & 0.140 & 0.082 & 0.73 & 0.73 \\ 
  $\widehat{\Delta}_{modsv}$ & -0.002 & 0.139 & 0.139 & 0.94 & 0.94 \\ 
  $\widehat{\Delta}_{modsv}^{cor}$ & 0.001 & 0.140 & 0.139 & 0.94 & 0.94 \\ 
  \hline
  $\widehat{\Delta}_{sv}^{only}$ & -0.001 & 0.139 & 0.140 & 0.94 & 0.94 \\ 
  $\widehat{\Delta}_{IPW}^{Bootstrap}$ & -0.001 & 0.139 & 0.139 & 0.94 & 0.94 \\ 
  $\widehat{\Delta}_{IPW}^{M-est}$ & -0.001 & 0.139 & 0.140 & 0.94 & 0.94 \\ 

   \hline
\end{tabular}
\caption{Simulation results for $(\alpha_0,\alpha_1)=(-3.76,4)$ and $\beta_3=-0.6$ under double-layer simulation set-up. See Section \ref{sec:sim_est} for a description of the estimators and Section \ref{sec:sim_setup} for a description of the summaries.}
\label{tb:4}
\end{table}

Bias column in Table \ref{tb:4} shows the difference of bias across all the estimators. The comparison between the PATE estimators is the same as that in Table \ref{tb:3}. Except that, IPWEs has almost the same bias as the model-based estimators with correct analysis model. In terms of the true variability of the PATE estimators, $\widehat{\Delta}_{OLS}^{cor}$ has the smallest MC SD as expected and the rest of the PATE estimators have similar variability. 

Ave SE of the model-based PATE estimators is computed with survey-based and model-based linear combination estimators; Ave SE of IPWEs is measured with survey-based linear combination method, parametric M-estimation, and the special bootstrap method. The Ave SE of all the model-based PATE estimators is very similar to their corresponding MC SD, except WOLS, which is same pattern as we see in the Table \ref{tb:3}. All the three variance methods for IPWEs perform resemblingly and approximately equal to the MC SD. The special bootstrap method estimates the variances for IPWEs well. Thus, we speculate that this bootstrap method may estimate the variances for the model based treatment effect estimators properly as well. However, we do not show the corresponding simulation results in this paper. 

Consistent with the single-layer construction, the finite and infinite population coverages are almost the same for the PATE estimators. Almost all of the coverage rates of the PATE estimators reach the nominal level, except WOLS. Note that $\widehat{\Delta}_{OLS}$ due to the large bias is not discussed here. 

The same comparative performance of the bias, Ave SE, and the coverage rates holds across the $\alpha$ and $\beta$ settings. Moreover, the trend of IPWEs shown in the Figure \ref{figure:4}is similar to the Figure \ref{figure:2}.   

\begin{figure}[H]
\centering
\includegraphics[width=0.4\textwidth]{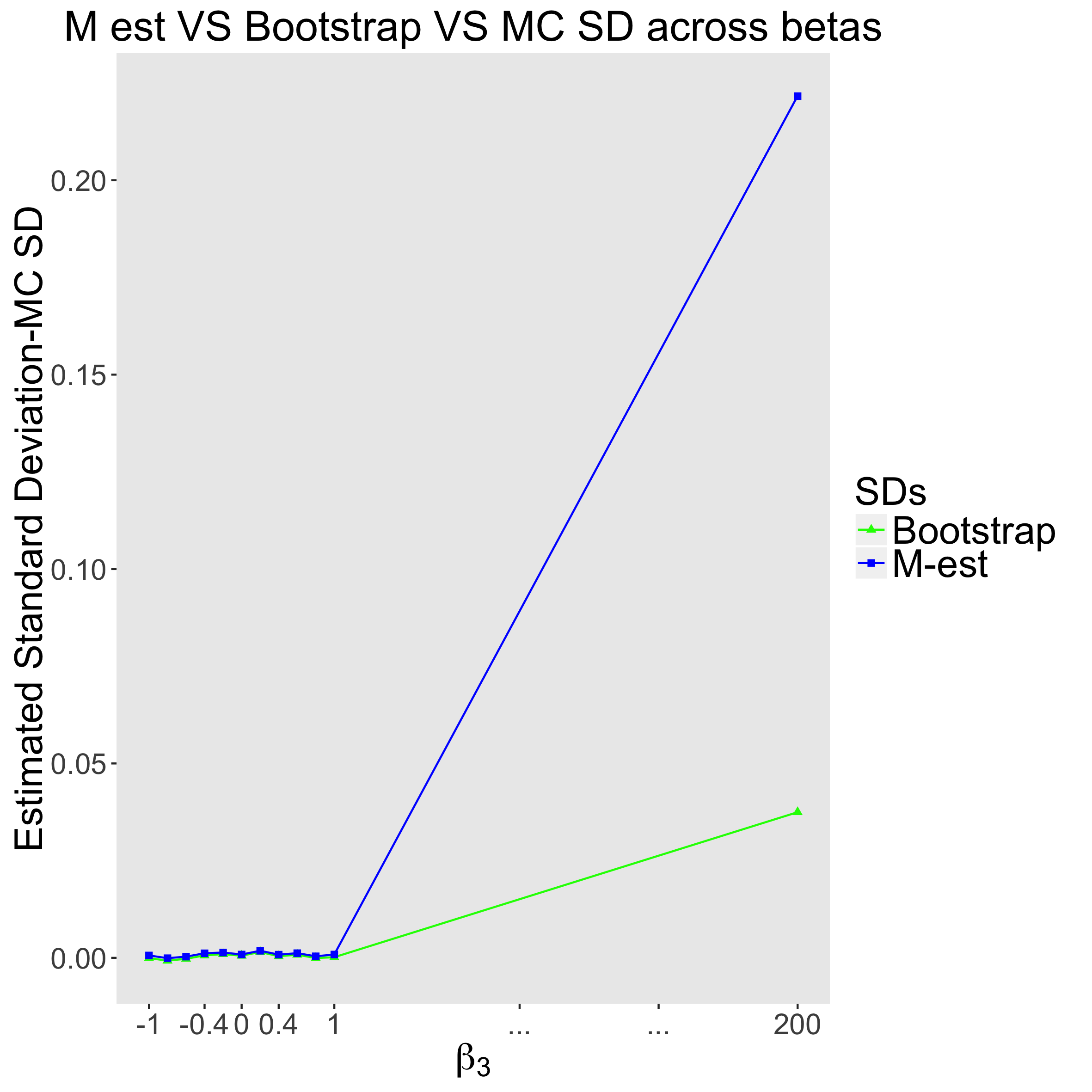}
\includegraphics[width=0.4\textwidth]{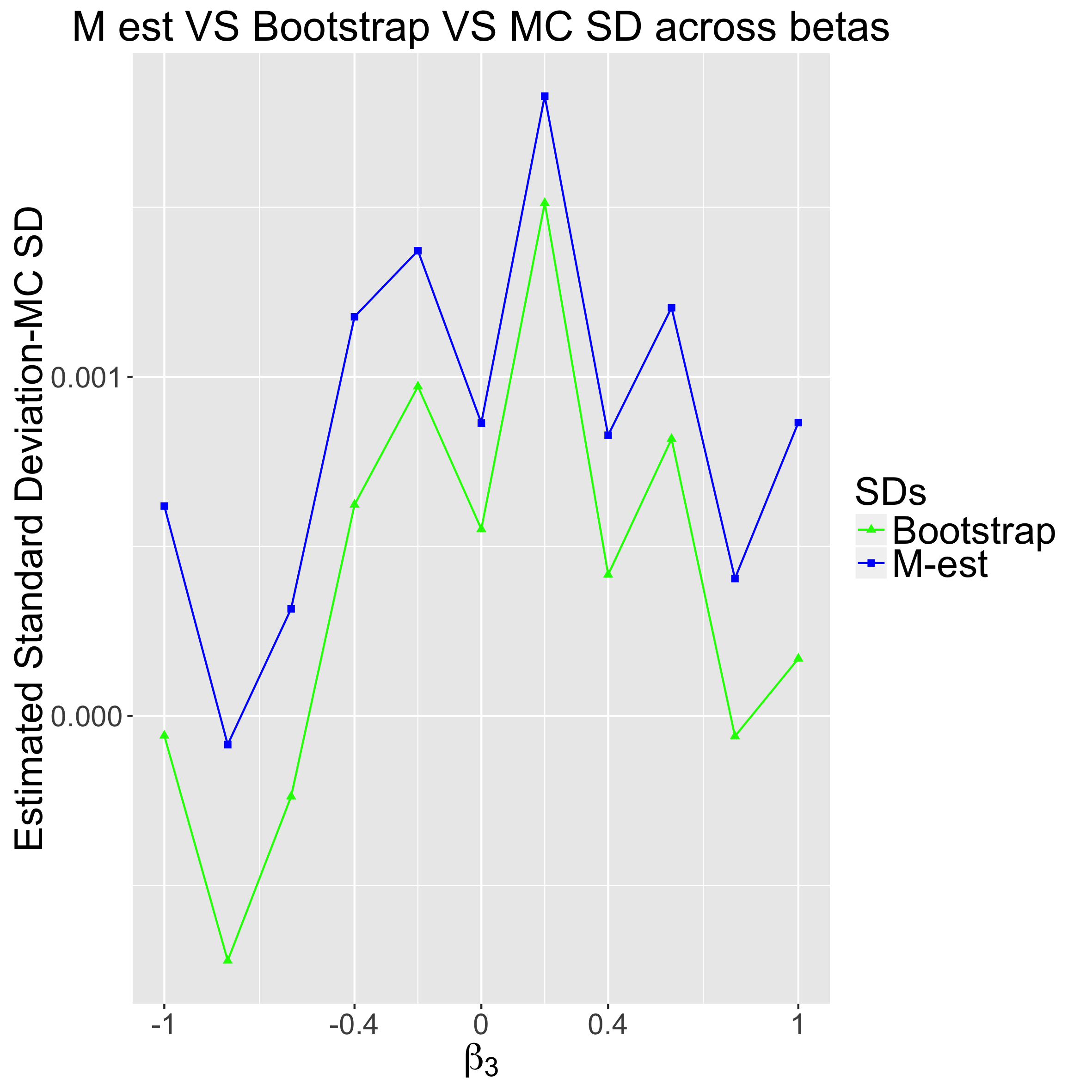}
\caption{The results of the bias of M-est and the special bootstrap method compared with MC SD for $(\alpha_0,\alpha_1)=(-3.76,4)$. The left panel covers the extreme case of $\beta_3$ and the right panel zooms in to the range of $\beta_3$, (-1,1).}
\label{figure:5}
\end{figure}

Across the settings in Table \ref{tb:2} with moderate treatment heterogeneity, the M-estimation standard deviation estimates are only slightly larger than the special bootstrap standard deviation estimates and the true SD, even though the M-estimation method is designed for accounting for all the sources of variability. As shown in Figure \ref{figure:5}, the blue line of M-estimation standard deviation fluctuates above 0 and the green line of the special bootstrap estimates. However, as $\beta_3$ increases dramatically, the estimates from the M-estimation method clearly exceeds the bootstrap estimates outstandingly and thus overestimates the standard deviation of IPWEs strikingly, as shown by the differences between the two types of lines of the left panel in Figure \ref{figure:5}.

 The trend regarding the two types of coverage rates in the double-layer set up is also identical to that in the single-layer set up. For $\alpha_1$ close to $0$, the $95 \%$ confidence interval for the IPWE captures both the infinite population average treatment and the specific target population average treatment nearly at the nominal level, as shown by the highest two lines in the center and right panels of Figure \ref{figure:4}. However, the coverage degrades for larger magnitudes of $\alpha_1$. The two types of coverage rates for the IPWEs are almost identical across all the scenarios as shown in Figure \ref{figure:4}. This trend is as we expected, since the large sample sizes would result in minimal differences between the finite- and infinite-population PATE.

\subsubsection{Study of Effect of Sample Size and Proportion}\label{sim_result_2}

We further conduct a second simulation study to examine the effect of sample size and proportion on the performance of the PATE estimators, especially IPWEs and the corresponding variance estimation methods. The variance estimation methods particularly heavily rely on the asymptotic environment as shown in Simulation Study 1. Thus, we follow the framework in the Simulation Study 1 to probe into the effect of sample size and proportion in both single-layer and double-layer constructions. 
\paragraph{Single-layer Simulation Results}\label{sim_result_2_inf}
Similar to Simulation Study 1, same measures are taken in each simulation with parameters as shown in Table \ref{tb:5}. We examine the performance of the PATE estimators across the settings 13-18 based on Bias and MC SD. The patterns of the performance among the PATE estimators are similar to that in Simulation Study 1, except that the scale of Bias and MC SD increases as $n_{total}$ decreases. Table \ref{tb:6} shows the Bias and MC SD of the PATE estimators under one setting 15. As exhibited in Simulation Study 1, under realistic situation without prior knowledge of the outcome analysis model, IPWEs outperform the model-based estimators with smaller bias and moderate variability. Thus, we concentrate on the performance of IPWEs, i.e. $\widehat{\Delta}_{sv}^{only}$ and $\widehat{\Delta}_{IPW}$,  and the corresponding variance estimation methods for the rest of this section. 

\begin{table}[H]
\centering
\begin{tabular}{c|cc}
  \hline
        Est& Bias &  MC SD  \\ 
  \hline
$\widehat{\Delta}_{OLS}$ & -0.286  & 0.324 \\ 
  $\widehat{\Delta}_{OLS}^{cor}$ & -0.008  & 0.578  \\ 
  $\widehat{\Delta}_{WOLS}$ & -0.079  & 0.519  \\ 
  $\widehat{\Delta}_{WOLS}^{cor}$ & -0.007  & 0.636  \\ 
  $\widehat{\Delta}_{modsv}$ & -0.079 & 0.519  \\ 
  $\widehat{\Delta}_{modsv}^{cor}$ & -0.007  & 0.636 \\ 
  $\widehat{\Delta}_{sv}^{only}$ & -0.042  & 0.552 \\ 
  $\widehat{\Delta}_{IPW}$ & -0.042  & 0.552 \\ 
  \hline
\end{tabular}
\caption{Simulation results for $(\alpha_0,\alpha_1)=(-3.76,4)$ and $n_{total}=200$ under setting 15 in Table \ref{tb:5}. }
\label{tb:6}
\end{table}

\begin{figure}[htp]
        \centering
        \begin{subfigure}[b]{0.475\textwidth}
            \includegraphics[height=4.5cm, left]{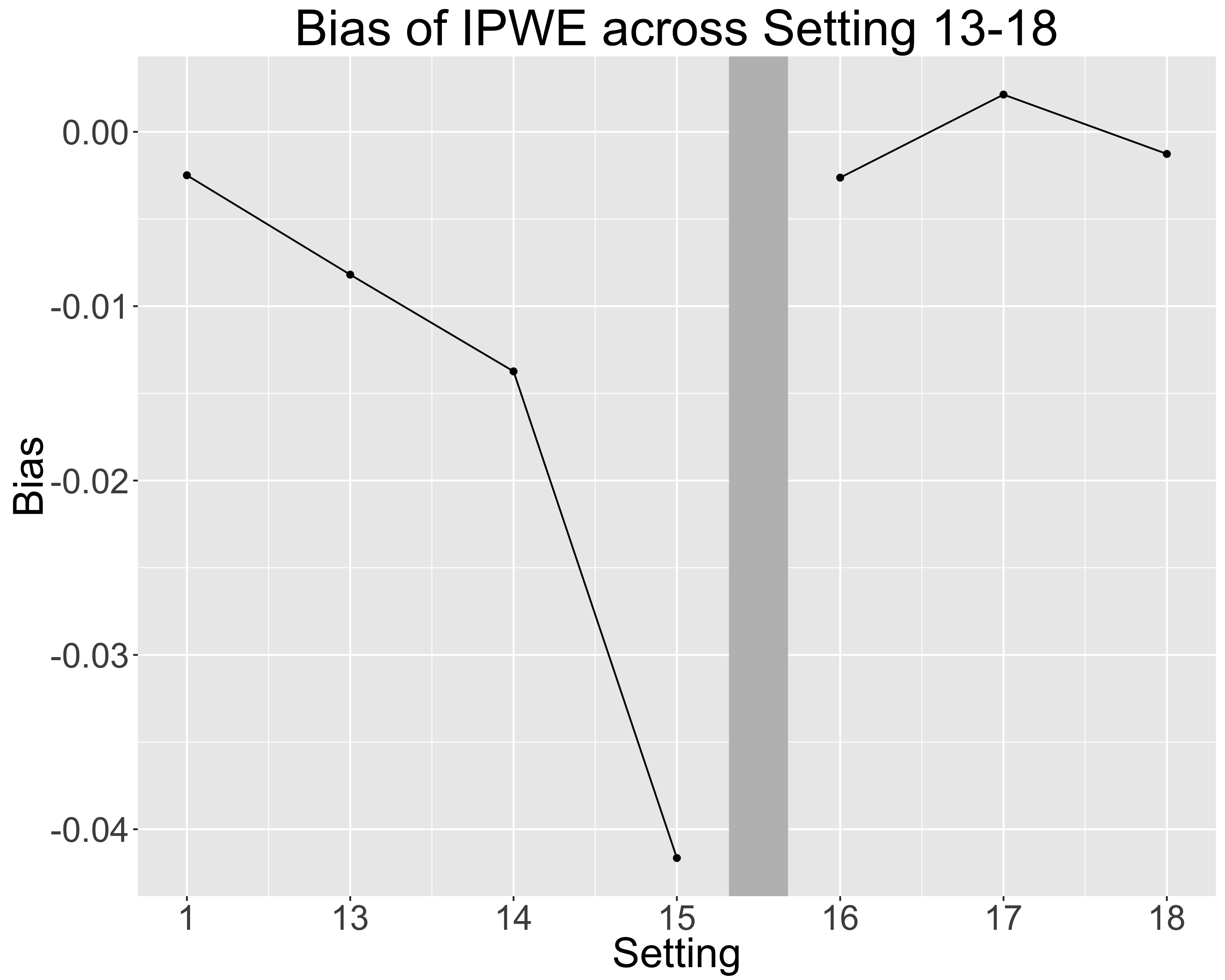}
        \end{subfigure}
        \hfill
        \begin{subfigure}[b]{0.475\textwidth}  
            \includegraphics[height=4.5cm, left]{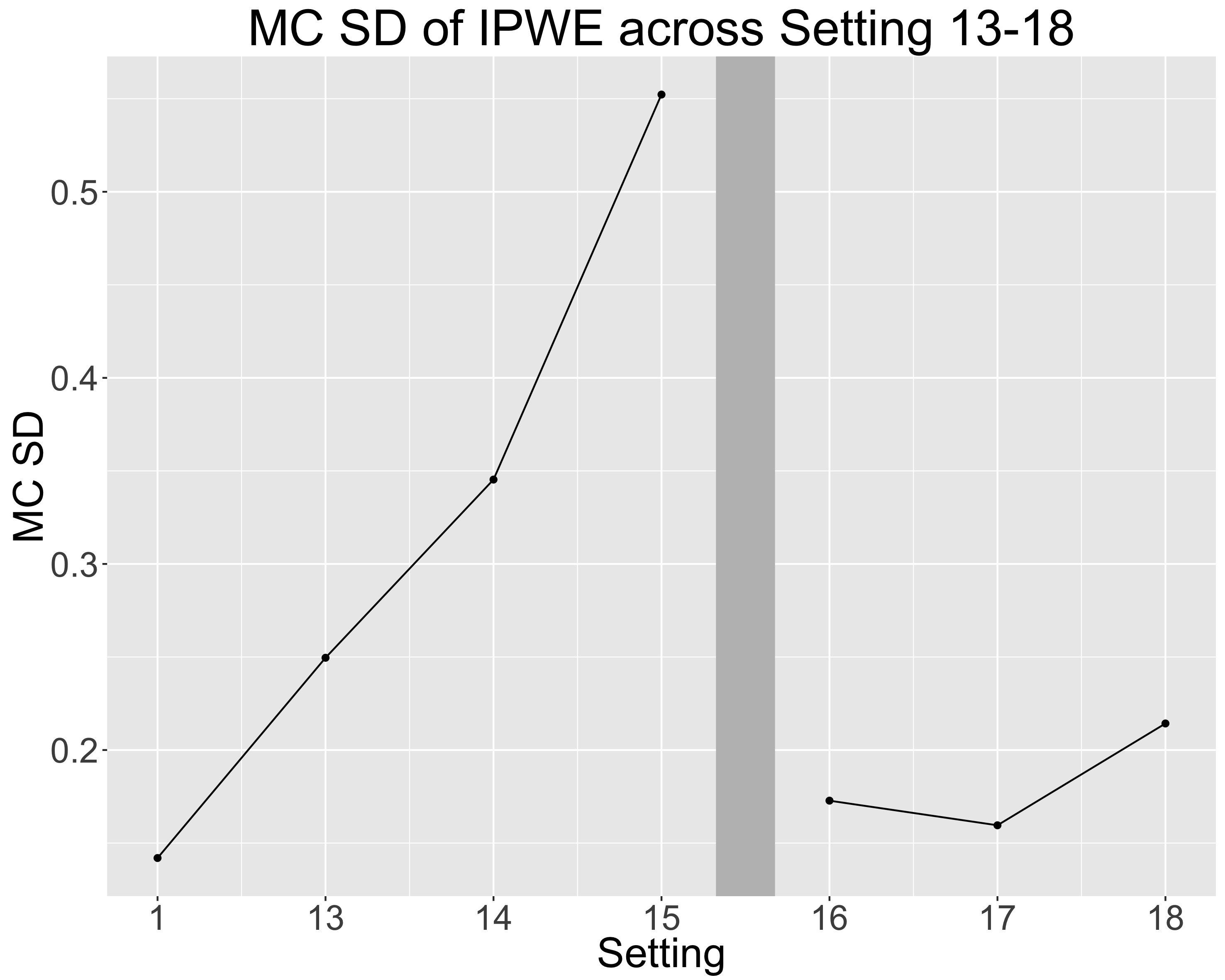}
        \end{subfigure}
        \vskip\baselineskip
        \begin{subfigure}[b]{0.475\textwidth}   
            \includegraphics[height=4.5cm, left]{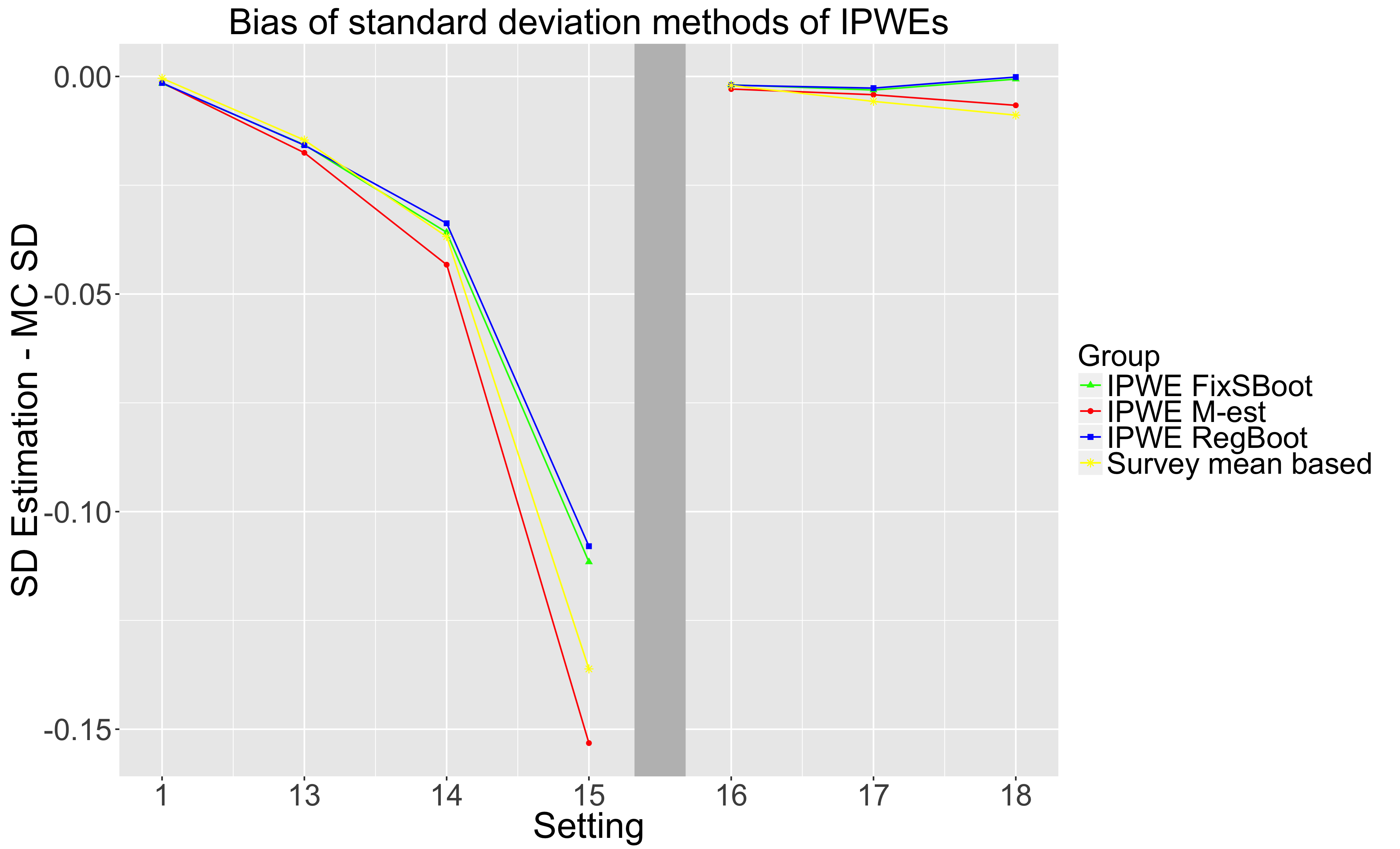}
        \end{subfigure}
        \quad
        \begin{subfigure}[b]{0.475\textwidth}   
            \includegraphics[height=4.5cm, left]{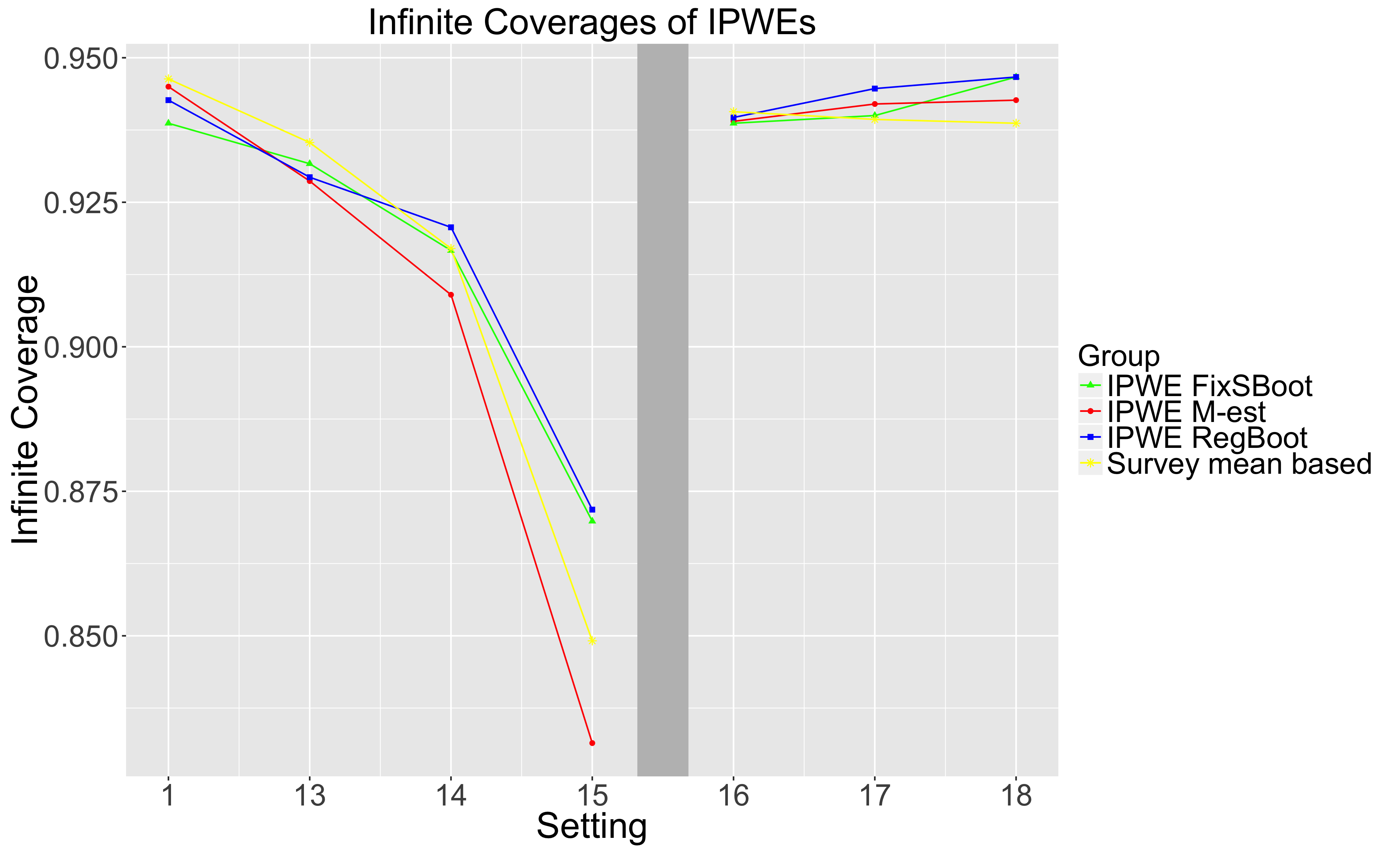}
        \end{subfigure}
        \caption[ The average and standard deviation of critical parameters ]
        {\small The first row of the panel is bias and MC SD of IPWEs; the second row of the panel is the bias of the standard deviation estimation methods, which include the M-estimation, Fixed S bootstrap, regular bootstrap, and the linear combination of the variances using survey sampling method and 95$\%$ infinite coverage rates of IPWE across simulation settings 1 and settings 13-18 under single-layer simulation set up. } 
        \label{figure:6}
    \end{figure}

The bias and the variability (MC SD) of IPWEs dramatically increase, as solely the combined size, $n_{total}$ decreases across settings 1,13,14, and 15 summarized in Table \ref{tb:5}, shown on the top panel in Figure \ref{figure:6}. As we shrink the size of $n_{total}$, we obtain less information from the trial to infer the PATE. Thus, the bias and the variability increases as well. However, as we increase the sample proportion from setting 16 to setting 18 summarized in Table \ref{tb:5}, we gain more information from the trial to infer the PATE  and thus the bias and the variability of IPWEs decrease nearly to the level before(setting 1). 


The variance estimation methods of $\widehat{\Delta}_{sv}^{only}$ and $\widehat{\Delta}_{IPW}$ including the parametric variance estimation methods and the bootstrap methods underestimate MC SD substantially as $n_{total}$ drops. The parametric variance method (M-estimation) of $\widehat{\Delta}_{IPW}$ deviates most, since it heavily relies on the asymptotic theories, while the bootstrap methods deviates least.  However, as the sample proportion climbs up, with more trial information retained, all the methods estimate closely to MC SD as shown in lower left panel of Figure \ref{figure:6}. The bootstrap methods, i.e. the regular bootstrap and the fixed S bootstrap of $\widehat{\Delta}_{sv}^{only}$ and $\widehat{\Delta}_{IPW}$ return similar estimates, so the lower left panel of Figure \ref{figure:6} only includes the bootstrap results of $\widehat{\Delta}_{IPW}$. 

The finite coverage rates and infinite coverage are almost the same from setting 13 to 18. The coverage rates declines as $n_{total}$ and then raise up to the nominal level as the sample proportion increases. 

\begin{figure}[htp]
        \centering
        \begin{subfigure}[b]{0.475\textwidth}
            \includegraphics[height=4.5cm, left]{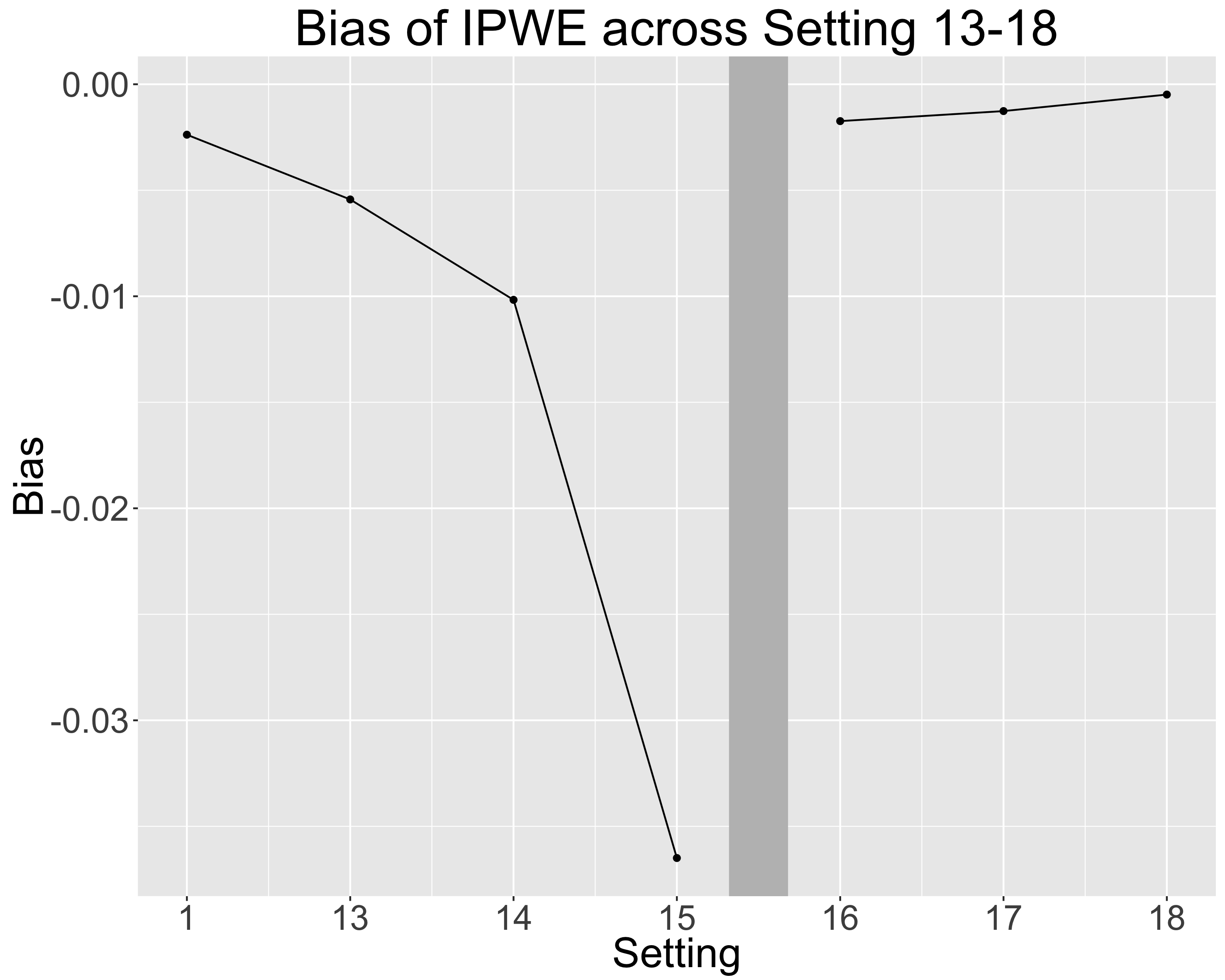}
        \end{subfigure}
        \hfill
        \begin{subfigure}[b]{0.475\textwidth}  
            \includegraphics[height=4.5cm, left]{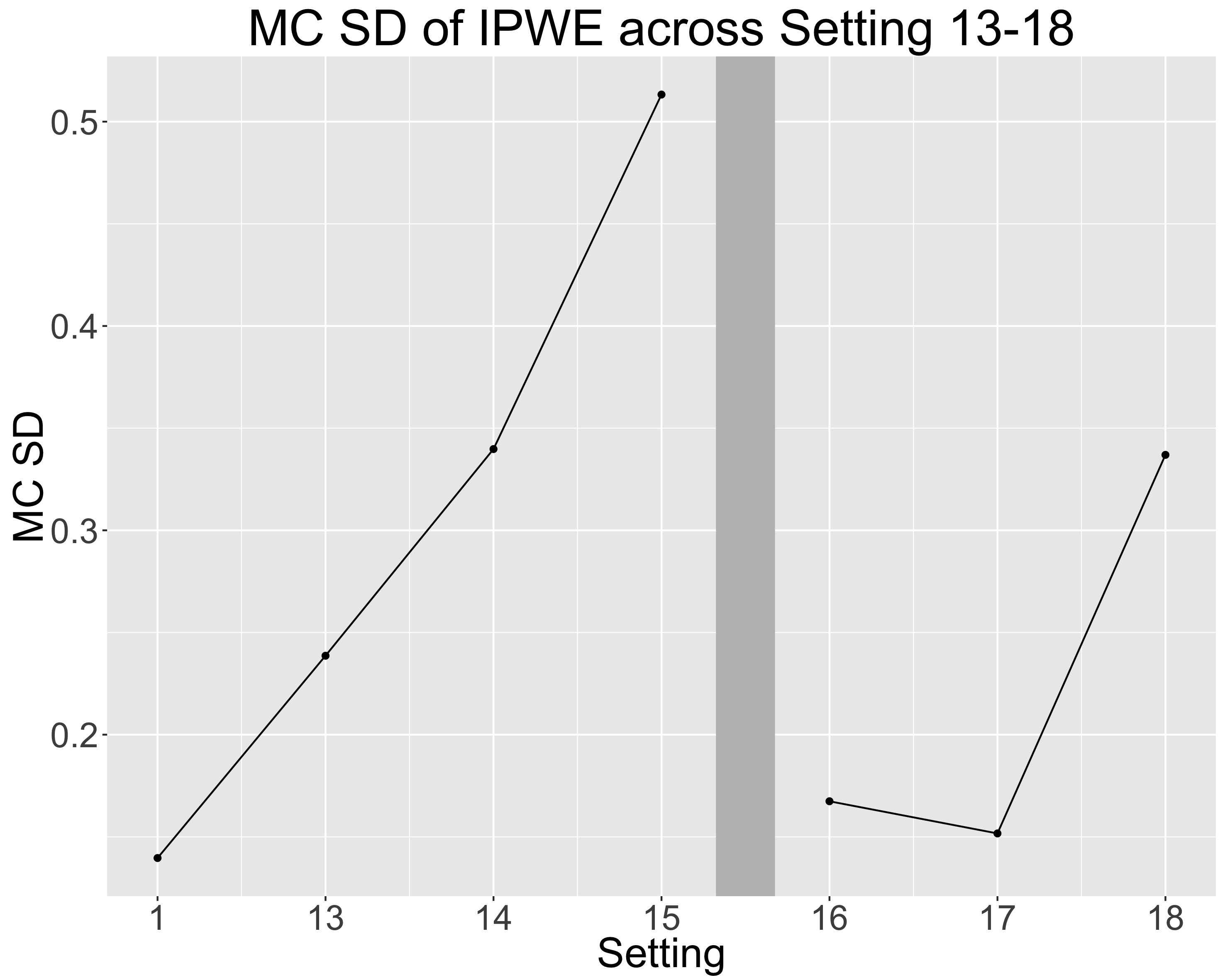}
        \end{subfigure}
        \vskip\baselineskip
        \begin{subfigure}[b]{0.475\textwidth}   
            \includegraphics[height=4.5cm, left]{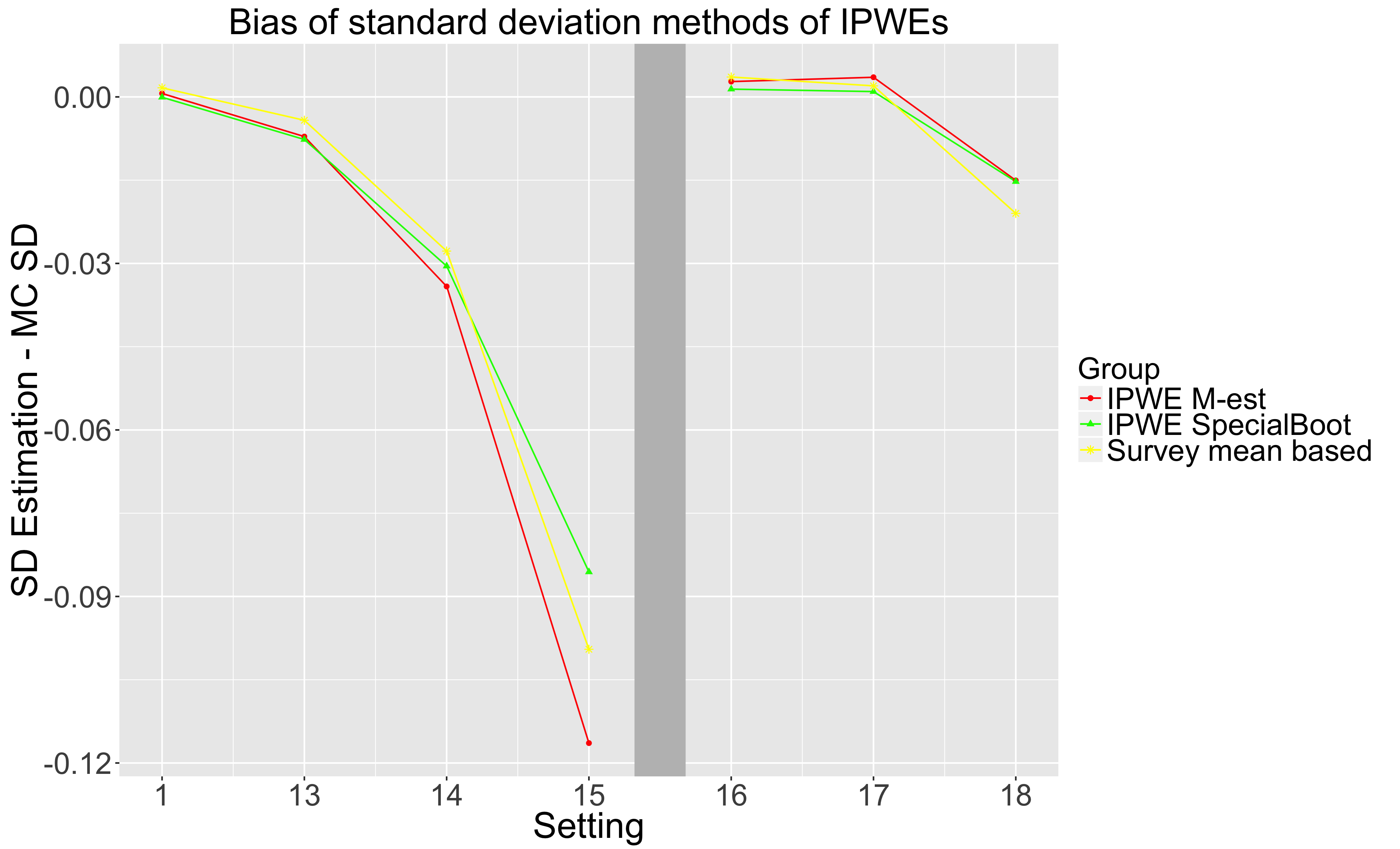}
        \end{subfigure}
        \quad
        \begin{subfigure}[b]{0.475\textwidth}   
            \includegraphics[height=4.5cm, left]{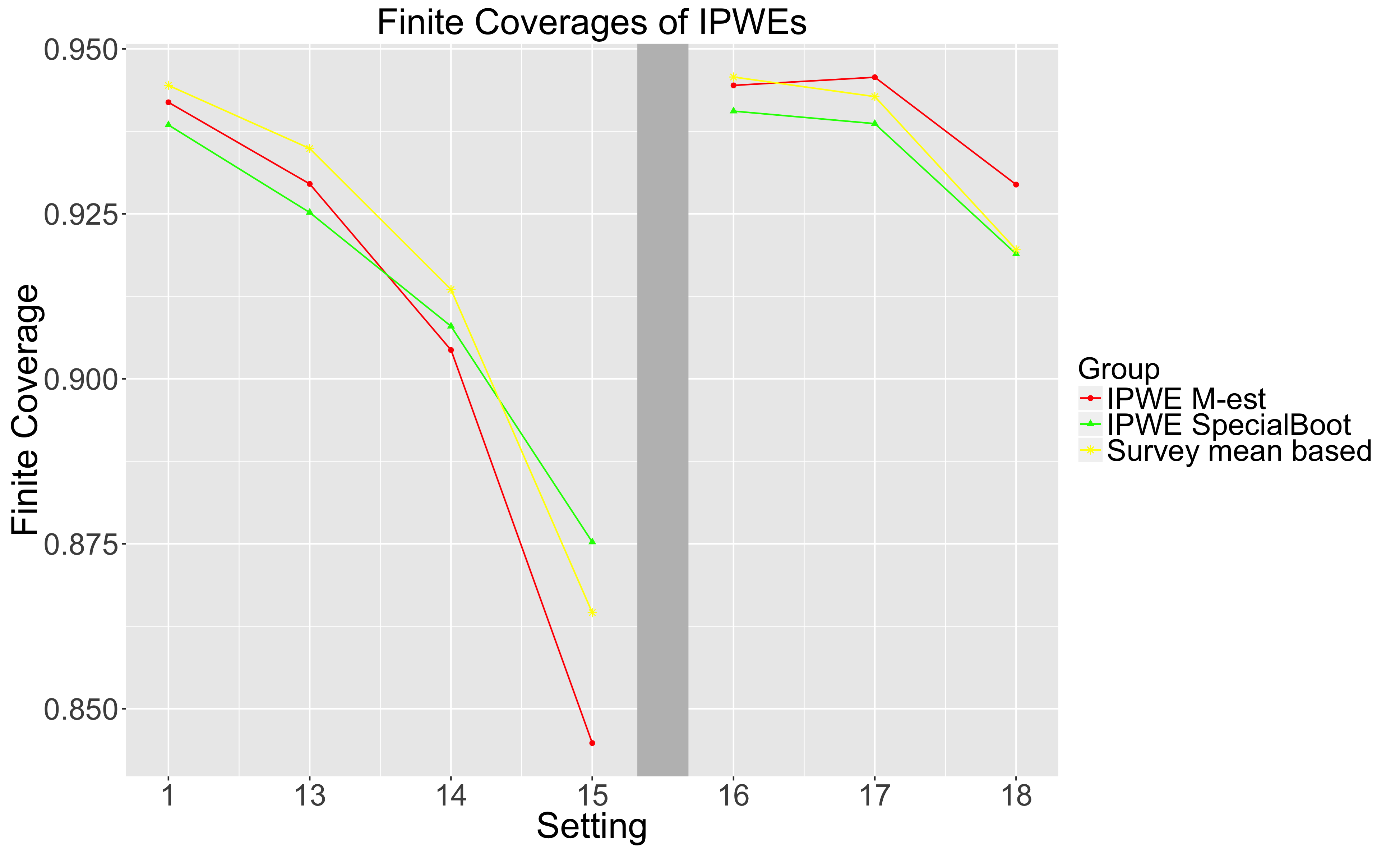}
        \end{subfigure}
        \caption[ The average and standard deviation of critical parameters ]
        {\small The first row of the panel is bias and MC SD of IPWEs; the second row of the panel is the bias of the standard deviation estimation methods, which include the M-estimation, the special bootstrap, and the linear combination of the variances using survey sampling method and 95$\%$ finite coverage rates of IPWE across simulation settings 1 and settings 13-18 under double-layer simulation set up. } 
        \label{figure:7}
    \end{figure}

\paragraph{Double-layer Simulation Results}\label{sim_result_2_fin}
The performance of IPWEs behave similarly as single-layer simulations. The decreasing $n_{total}$ drags down the performance of IPWEs, but the increasing sample trial improves the performance of IPWEs as shown in top panel of Figure \ref{figure:7}. 



The variance estimation methods we consider here are the parametric variance estimation method of $\widehat{\Delta}_{sv}^{only}$, the special bootstrap for $\widehat{\Delta}_{IPW}$ and its M-estimation method. Those methods again insufficiently estimate the variability of IPWEs with $n_{total}$ small, but the special bootstrap and the parametric variance estimation method of $\widehat{\Delta}_{sv}^{only}$ are more reliable than the M-estimation, which deviates the most as shown in lower left panel of Figure \ref{figure:7}. With large proportion of sample size all the variance methods estimate the variability well. Finite coverage rates are still almost the same as infinite coverage rates. Thus, Figure \ref{figure:7} only shows the finite coverage rates and the trend is similar to those in single-layer simulation. 

IPWEs perform poorly as $n_{total}$ decreases, especially the variance estimation methods. The performance of the M-estimation and the bootstrap methods depend on the asymptotic theories. Thus small $n_{total}$ breaks and drags these methods down. However, when there are relatively large size of trials, the variance estimation recovers from the small size of $n_{total}$ and thus, perform well. In terms of the simulations, as $n_{total}$ decreases, we drop several MC data sets due to insufficient target population size.

\section{Empirical Example: Prevention and Treatment of Hypertension Study}\label{sec:ex}
In this section, we demonstrate the use of IPWE with variances estimated via M-estimation  and the special bootstrap method on a real empirical example. We generalize the results of a randomized trial of an alcohol moderation treatment on blood pressure among moderate or heavy alcohol veteran drinkers with hypertension to this subset of the U.S. population. We obtain trial data from Prevention and Treatment of Hypertension Study (PATHS) sponsored by the National Heart, Lung, and Blood Institute (NHLBI). The detailed design, inclusion criteria, and the trial original goal can be found at https://biolincc.nhlbi.nih.gov/studies/paths/. We extract the population data from the National Health and Nutrition Examination Survey (NHANES) at https://wwwn.cdc.gov/nchs/nhanes/. 

\begin{table}[H]
\centering
\begin{tabular}{c|c|c}
  \hline
& \begin{tabular}{@{}c@{}}PATHS \\ $(size=616)$\end{tabular} & \begin{tabular}{@{}c@{}}NHANES \\ $(size=83)$\end{tabular} \\ 
 \hline
 \begin{tabular}{@{}c@{}}Age, year\\ $(SD)$\end{tabular}& \begin{tabular}{@{}c@{}}58.9\\ $(10.6)$\end{tabular}& \begin{tabular}{@{}c@{}}53.2\\ $(13.4)$\end{tabular}\\
 \begin{tabular}{@{}c@{}}Baseline DBP, mmHG\\ $(SD)$\end{tabular}& \begin{tabular}{@{}c@{}}86.6\\ $(5.2)$\end{tabular}&\begin{tabular}{@{}c@{}}81.9\\ $(7.0)$\end{tabular}\\

 \hline
\end{tabular} 
\caption{Baseline characteristics in PATHS and NHANES. Values are mean (SD). }
\label{tb:6}
\end{table}

The PATHS trial in total includes 616 subjects, recruited from 1989-1994. The subjects were randomized either to an alcohol reduction treatment or to a control condition. Based on Cushman, \textit{et al}'s \cite{cushman1998} literature review, the statistically significant treatment modifiers affecting the change of DBP are age and the baseline measure of DBP before the treatment, even though additional potential treatment modifiers are included in the PATHS, e.g., body mass, sodium and potassium excretion, weight, and so on (\cite{cushman1998, xin2001}). While the study of Cushman \textit{et al.} (1998) \cite{cushman1998} considers several outcome measures, e.g. systolic blood pressure and diastolic blood pressure(DBP), we only focus on one of the main problems the study discussed, whether DBP is reduced in the 6 months after the treatment. Table 5 of Cushman\textit{et al.}'s study (1998) \cite{cushman1998} shows that the intervention group had a 0.3 mm Hg greater reduction in DBP at the 6th month compared with baseline DBP measures with standard error 0.6 mm Hg controlling for the confounding factors mentioned above. Thus, on average, there was little evidence of an effect of intervention. However, because of potential treatment heterogeneity, a significant difference could appear in a different population. 

The population data we select is obtained from NHANES 1999-2000 Questionnaire Data, the year nearest to the study year of PATHS. We subset the population data based on exactly same inclusion and exclusion criteria of PATHS study to ensure the similarity between the trial and the target population. We end up with only 83 subjects. These 83 population subjects can be considered as either the specific finite target population or a target population randomly drawn from the infinite population. For the purpose of demonstrating our method, we consider the finite population to be the outpatient veteran survey participants who reported moderate or heavy alcohol intake per day in $1999-2000$. The corresponding infinite population can be defined as all possible moderate or heavy veteran drinkers across U.S.with characteristics similar to the survey participants. 
Table \ref{tb:6} summarizes baseline characteristics in PATHS and NHANES. 

\begin{figure}[H]
\centering
\includegraphics[width=0.5	\textwidth]{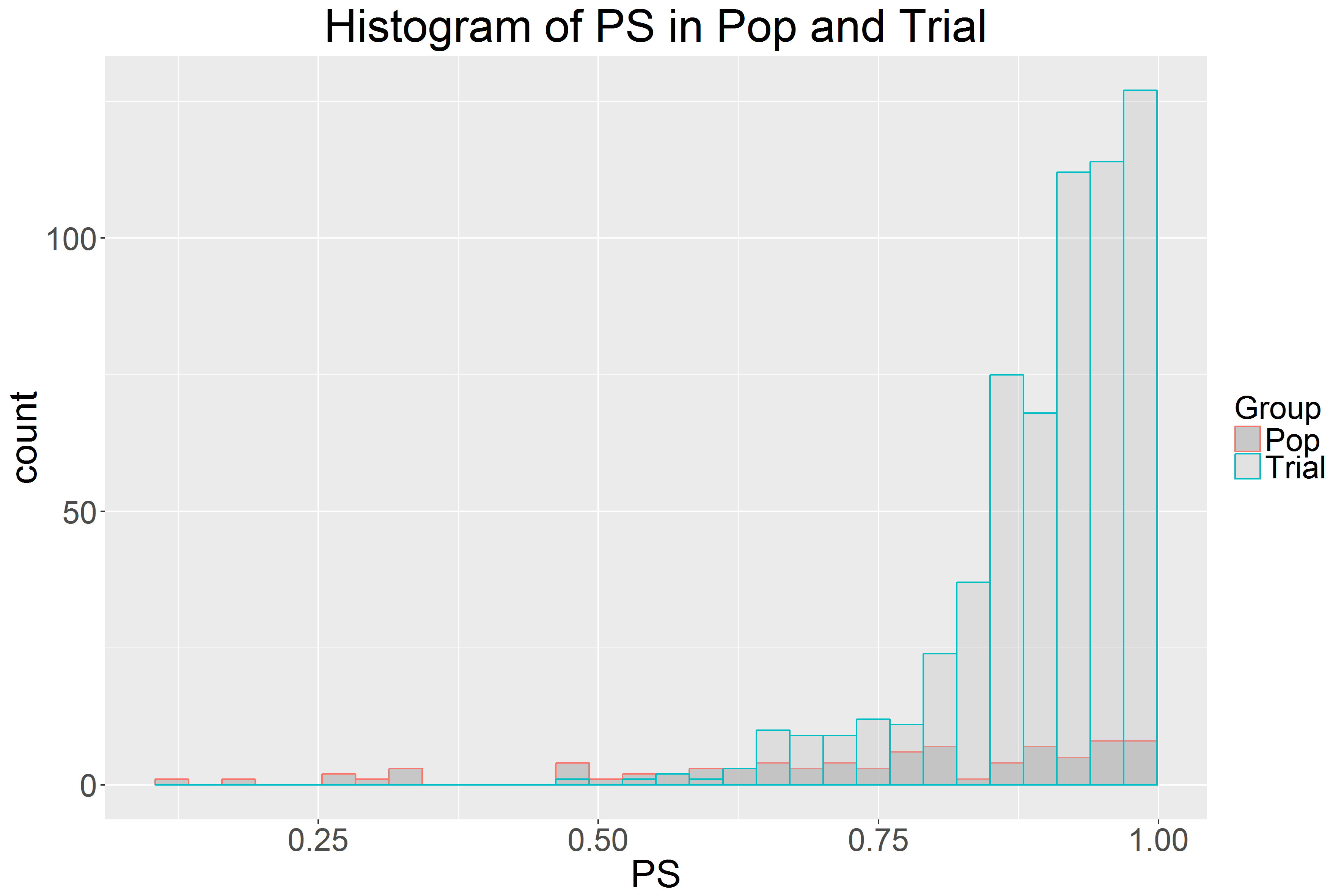}
\caption{The distributions of PS in the trial and the target population. }
\label{figure:ex1}
\end{figure}
We compute $\Delta_p$ to measure the similarity of the covariate distribution between the target population and the trial. Figure \ref{figure:ex1} exhibits the distributions of PS in the trial and the target population, with $\Delta_p=0.164$. There is considerable  overlap between the distribution of the target and the population. 

\begin{table}[H]
\centering
\begin{tabular}{c|ccc}
  \hline
  &Estimate &SE&$95 \%$ CI (mm HG)\\
  \hline
 RCT (PATHS) &0.300&0.600& (-0.88,1.48)\\
 Infinite pop M-est &0.188&0.807& (-1.39, 1.77)\\
Finite pop bootstrap  &0.188&0.756&(-1.29, 1.67)\\
 \hline
\end{tabular} 
\caption{Baseline characteristics in PATHS and NHANES. Values are mean (SD). }
\label{tb:7}
\end{table}

We apply IPWE and the corresponding variance estimation methods, M-estimation and the special bootstrap to generalize the trial results of PATHS (as shown in the first line in Table \ref{tb:7}) to the target population, NHANES. Although NHANES is a complex sample survey, we chose not to incorporate the survey design in the weight estimation, as this is beyond the scope of our proposed and comparative methods. 
We calculate $\widehat{\Delta}_{IPW}$ as 0.188 mm HG of DBP with estimated standard deviation using M-estimation method, 0.807 mm HG. If we are interested in the average treatment effect for all possible moderate or heavy veteran drinkers across U.S. and also across years, the 95 $\%$ confidence interval for the average treatment effect is -1.39 to 1.77 mm HG. Thus, the alcohol reduction treatment may not be effective on hypertension among U.S. veterans. We apply the special finite population bootstrap method to estimate the standard deviation of $\widehat{\Delta}_{IPW}$, 0.756 mm HG, which is slightly smaller than the results of the M-estimation method. If we only concentrate on 1999 U.S. moderate or heavy veteran drinkers, the confidence interval for the average treatment effect is between -1.29 and 1.67 mm HG. The results are summarized in Table \ref{tb:7}. 

We also calculate the model-based estimators described in Section \ref{sec:method_est} and \ref{sec:sim_est}, including OLS, WOLS, and survey based estimators to compare with IPWE. For those model-based estimators, we choose to use a linear combination of variances method to estimate the standard deviation. We decide the outcome analysis model as 
\begin{equation}
Y=\rho_0+\rho_1Age+\rho_2BaselineDBP+\rho_3Trt+\rho_4Trt*BaselineDBP+\epsilon
\end{equation}
$Y$ is the change of DBP at 6th month compared with baseline DBP and $\epsilon$ represents the independent normal random errors. We do not add another interaction term between Trt and Age due to its extreme small effect size. Then the model-based PATE estimators shares the form, $\hat{\rho_3}+\hat{\rho_4} \overline{BaselineDBP}_{pop}$. The results are summarized in Table \ref{tb:7}.

\begin{table}[H]
\centering
\begin{tabular}{c|ccc}
  \hline
  &Estimate &SE&$95 \%$ CI (mm HG)\\
  \hline
 $\widehat{\Delta}_{WOLS}$ &0.512&0.618& (-0.70, 1.72)\\
$\widehat{\Delta}_{modsv}$ &0.512&0.931&(-1.31, 2.34)\\
$\widehat{\Delta}_{sv}^{only}$  &0.188&0.806&(-1.39, 1.77)\\
 \hline
\end{tabular} 
\caption{Baseline characteristics in PATHS and NHANES. Values are mean (SD). }
\label{tb:7}
\end{table}

Compared with the model-based estimators, we recommend to choose IPWEs to generalize the trial results to the target population. The results we have in Table \ref{tb:7} exhibit a similar pattern in Table \ref{tb:3} and \ref{tb:4}. Compared with our simulation scenarios, $\Delta_p$ lies between the second ($\alpha_1=0$) and the third scenario ($\alpha_1=4$) in Table \ref{tb:1} and much closer to the second scenario with $\Delta_p=0.15$. $\widehat{\Delta}_{OLS}$ presumably should have largest bias without prior knowledge of the analysis model and in Table \ref{tb:7} it has inconsistent estimates compare with other estimators. $\widehat{\Delta}_{WOLS}$ and $\widehat{\Delta}_{modsv}$ with same analysis model have similar estimates for PATE, while the estimated standard deviation of survey based estimator is larger than WOLS as expected. $\widehat{\Delta}_{sv}^{only}$ and  $\widehat{\Delta}_{IPW}$ have similar estimated standard deviation. Without any knowledge of the outcome analysis model, IPWEs are the safe choice to use. Furthermore with large sample proportion, approximately $88 \%$ and moderate size of $n_{total}=699$, we have more confidence to prefer $\widehat{\Delta}_{sv}^{only}$ and $\widehat{\Delta}_{IPW}$ to the other model-based estimators for the two types of the target population.  

The generalization to the target population is necessary if the researchers try to implement the alcohol moderation treatment in PATHS to some target population, even though the bias of IPWEs might be a concern. The plot \ref{figure:ex1} clearly shows that the distributions of PSs are quiet different between the PATHS trial and the target population. Thus, with moderate size of $\Delta_p$, the results of PATHS need to be tuned to suit for the target population. However, the distribution of the target population is more left skewed than that of the trial. The non-overlapping part might arise the concern of the bias of the IPWEs, but based on the simulation studies we conduct, IPWEs are suggested to use and they should have promising coverage rates of the true PATE.

 \section{Discussion}\label{sec:discuss}
In our study, we concentrated on reweighing methods, first carefully defining the parameter of interest, then studying different model-based average treatment effect estimators and IPWEs to weight the trial results to represent the target population, and proposing several variance estimation methods for IPWEs. In particular, we proposed a new variance estimator and develop several bootstrap-based variance estimation methods for IPWEs under different types of target populations. We conducted two simulation studies to examine the properties of the PATE estimators and the proposed variance estimators under various magnitudes of selection and heterogeneity effect sizes and different sample size situations. We also demonstrated these methods in a realistic data example. 

We define PATE for different types of populations, the infinite and finite target population. We apply single-layer simulation construction to assess the properties of the PATE estimators and their corresponding variance estimation for the infinite target population and use double-layer simulation construction for the finite target population. The trend of the performances of the PATE estimators in the single-layer simulation construction is similar to that in the double-layer simulation construction. The variance estimation methods behave differently in these two simulation constructions, which is summarized later. 

Our study can provide guidance to practitioners as to which PATE estimator to use in specific data situations. The performance of the PATE estimators depends on the similarity between the trial and the target populations, the size of the heterogeneity of the treatment effect, the size of the populations, and the prior knowledge of the analysis model. If the target population covariate distribution is completely identical with the trial's, then weighted or unweighted, model-based or IPWE all perform well. It is not necessary to weight the trial results for a target population and it is safe to apply the trial results directly to the target population. As there is less overlap of the covariate distribution between the target population and the trial, i.e., $\Delta_p$ increases up to 0.38, then we concern the generalization results through any weighted PATE estimator due to large bias and low coverage rates. With $\Delta_p$ less than 0.38, IPWEs perform similarly to the other model-based estimators with similar bias. With low $n_{total}$, IPWEs outperform the other model-based estimators with smaller bias. As $n_{total}$ drops between 200 and 500, IPWEs are concerned to use due to low coverage rates. With high sample proportion, IPWEs can perform as well as if $n_{total}$ is large. However, even under the case of high sample proportion, i.e., $95\%$, with low $n_{total}$, i.e., between 200 and 500, IPWEs should be used with caution due to slightly low coverage rates. With $n_{total} \leq 200$, IPWEs are not suggested due to unstable computation results. 

Different variance estimation methods are suggested for the infinite and the finite target population. If we deal with the infinite target population, as $\Delta_p$ is between 0 and 0.15, the parametric variance estimation method, the regular bootstrap, and the fixed S bootstrap all provide similar  results closer to the true variability for all the PATE estimators, except the weighted OLS. For the weighted OLS, the boostrap methods outperform the parametric method. As $\Delta_p$ increases up to 0.38, all the variance estimation methods underestimate the true variability. If the target population is finite, as $\Delta_p$ is between 0 and 0.15, the parametric variance estimation method and the special bootstrap both provide similar  results closer to the true variability for the corresponding PATE estimators, except the weighted OLS, for which the boostrap method is recommended. As $\Delta_p$ increases up to 0.38, all the variance estimation methods underestimate the true variability. Especially for IPWEs, with extreme large size of the heterogeneity of the treatment effect, the special bootstrap provides better estimates than the M-estimation. 

The sample size and sample proportions affect the performance of variance estimation methods. No matter the types of the target population, as $n_{total}$ decreases between 200 and 500, all the variance estimation methods severely underestimate the true variability of the PATE estimators. Associated with the increasing sample proportion to $95 \%$, the variance estimation methods provide better estimates, but still the parametric method does not estimate the variability of the weighted OLS well. Especially for IPWEs, the bootstrap methods are always recommended. The M-estimation breaks easily as $n_{total}$ drops. Even with high proportions, i.e., $95\%$, if combined with low $n_{total}$, i.e., lower than 500, we should be cautious with the performance of the variance estimation methods.  

Now we can combine the findings of the PATE estimators with the performance of the corresponding variance estimation methods. Under moderate overlap of the covariate distributions between the trial and the target population, i.e. $\Delta_p \leq 0.15$ and large sample size, if the target population is infinite, the weighted OLS with bootstrap methods and the survey based weighted estimators with either the parametric method or the boostrap method perform as well as IPWE with either bootstrap methods or the parametric variance estimation method. All of them have smaller bias and the coverage rates closer to the nominal level. Under same scenarios, if the target is finite, then the survey based weighted estimators with the parametric variance estimation method and IPWEs with either the parametric methods or the bootstrap methods outperform than the others. Under other extreme cases, IPWEs with either bootstrap methods ($n_{total}$ is low) or the parametric methods outperform than the others with smaller bias and higher coverage rates.  

Limitations of the IPWEs and the proposed variance methods appear under the scenarios of extreme dissimilarity between the populations or small trial size, and of extreme large heterogeneity of treatment effect. First, when there is considerable lack of overlap between the covariate distribution of the target population and the trial (large $\Delta_p$) or there is a small trial, the model-based treatment effect estimators and IPWEs cannot recover the true PATE from the trial results successfully. Under those situations, without prior knowledge of the outcome analysis model, the PATE estimators have large bias and low coverage rates for the PATE regardless of the types of the target population and the variance estimation methods. Since the trial cannot represent some part of the target population, the weighted treatment effect estimators perform poorly. The variance estimation methods including the M-estimation and the bootstrap methods heavily rely on the asymptotic theories which require large sample size. When the sample size shrinks, those variance estimation methods collapse. 

Second, with large heterogeneity of treatment effect the model-based treatment effect estimators and IPWEs also perform poorly with large bias. But, since the methods overestimate the variance, IPWEs end up with higher coverage rates for the true PATE regardless of the types of the target populations. However, even though under these extreme scenarios, compared with the model-based treatment effect estimators, IPWEs still outperform them. 

\newpage 

\bibliography{Manuscript1} 
\bibliographystyle{plain}

\end{document}